\documentclass[12pt,a4paper]{article}

\usepackage[english]{babel}
\usepackage{amsmath,amsthm,amsfonts,bbm,graphicx,psfrag,cite}

\renewcommand{\Re}{\mathrm{Re}}

\newcommand{\Id}{\mathbbm{1}}
\newcommand{\Or}{\mathcal{O}}
\newcommand{\Z}{\mathbbm{Z}}

\newcommand{\R}{\mathbbm{R}}
\newcommand{\Pb}{\mathbbm{P}}
\newcommand{\dx}{\mathrm{d}}
\newcommand{\Ai}{\mathrm{Ai}}
\newcommand{\e}{\varepsilon}

\newcommand{\Dt}[2]{\frac{\dx #1}{\dx #2}}
\newcommand{\Af}{{\cal A}_{\rm 1}}
\newcommand{\Ac}{{\cal A}_{\rm 2}}
\newcommand{\I}{{\rm i}}
\newcommand{\cte}{{\rm const}\,}

\numberwithin{equation}{section}

\newtheorem{prop}{Proposition}[section]
\newtheorem{thm}[prop]{Theorem}
\newtheorem{lem}[prop]{Lemma}
\newtheorem{defin}[prop]{Definition}

\newenvironment{proofOF}[2]{\removelastskip\vspace{6pt}\noindent {\it Proof of #1.}~\rm#2}{\qed \par\vspace{6pt}}

\numberwithin{equation}{section}

\title{Large time asymptotics of growth models\\ on space-like paths I: PushASEP}
\author{Alexei Borodin\thanks{California Institute of Technology, Mathematics 253-37, Pasadena, CA 91125, USA. E-mail: borodin@caltech.edu.},
Patrik L. Ferrari\thanks{Weierstrass Institute for Applied Analysis and Stochastics, Mohrenstrasse 39, D-10117 Berlin, Germany. E-mail: ferrari@wias-berlin.de.}}

\date{13. August 2008}

\begin{document}
\maketitle \sloppy

\begin{abstract}
We consider a new interacting particle system on the one-dimensional
lattice that interpolates between TASEP and Toom's model:
A particle cannot jump to the right if the neighboring site is
occupied, and when jumping to the left it simply pushes
all the neighbors that block its way.

We prove that for flat and step initial conditions,
the large time fluctuations of the height function of the
associated growth model along any space-like path are
described by the Airy$_1$ and Airy$_2$ processes. This includes
fluctuations of the height profile for a fixed time and
fluctuations of a tagged particle's trajectory as special
cases.
\end{abstract}

\section{Introduction}\label{sectIntroduction}
We consider a model of interacting particle systems, which is a generalization of the TASEP (totally asymmetric simple exclusion process) and the Toom model. Besides the extension of some universality results to a new model, the main feature of this paper is the extension of the range of analysis to any ``space-like'' paths in space-time, whose extreme cases are fixed time and fixed particle (tagged particle problem), see below for details.

Consider the system of $N$ particles $x_1>\dots>x_N$ in $\mathbb{Z}$
that undergoes the following continuous time Markovian evolution:
Each particle has two exponential clocks -- one is responsible for
its jumps to the left while the other one is responsible for its
jumps to the right. All $2N$ clocks are independent, and the rates
of all left clocks are equal to $L$ while the rates of all right
clocks are equal to $R$. When the $i$th left clock rings, the $i$th
particle jumps to the nearest vacant site on its left. When the
$i$th right clock rings, the $i$th particle jumps to the right by
one provided that the site $x_i+1$ is empty; otherwise it stays put.
The main goal of the paper is to study the asymptotic properties of
this system when the number of particles and the evolution time
become large.

If $L=0$ then the dynamics is known under the name of Totally
Asymmetric Simple Exclusion Process (TASEP), and if $R=0$ the
dynamics is a special case of Toom's model studied in~\cite{DLSS91} (see references therein too). Both
systems belong to the Kardar-Parisi-Zhang (KPZ) universality class
of growth models in $1+1$ dimensions.

Particle's jump to the nearest vacant spot on its left can be also
viewed as the particle pushing all its left neighbors by one if they
prevent it from jumping to the left. This point of view is
often beneficial because it remains meaningful for infinite systems,
and also the order of particles is not being changed. Because of
this pushing effect we call our system the Pushing Asymmetric Simple
Exclusion Process or PushASEP.

Observe that for a $N$-particle PushASEP with particles \mbox{$x_1(t)>\dots>x_N(t)$}, the evolution of $(x_1,\dots,x_M)$ for any
$M\le N$ is the $M$-particle PushASEP not influenced by the presence
of the remaining $N-M$ particles. This "triangularity property"
seems to be a key feature of our model that allows our analysis to
go through.

Our results split in two groups -- algebraic and analytic.

Algebraically, we derive a determinantal formula for the
distribution of the $N$-particle PushASEP with an arbitrary fixed
initial condition, and we also represent this distribution as a gap
probability for a (possibly, signed) determinantal point process
(see~\cite{Lyo03,BKPV05,Sos06,Jo05,Spo05} for information on determinantal processes). The
result is obtained in greater generality with jump rates $L$ and $R$
being both time and particle-dependent (Proposition~\ref{PropPushASEP}). The first part
(the determinantal formula, see Proposition~\ref{Prop1}) is a generalization of similar results
due to \cite{Sch97,RS06,AKK99} obtained by the Bethe Ansatz techniques. Also, a closely related result have been obtained very recently in~\cite{DW07} using a version of the Robinson-Schensted-Knuth correspondence.

Analytically, we use the above-mentioned determinantal process to
study the large time behavior of the infinite-particle PushASEP with
two initial conditions:\\[0.5em]
1. Flat initial condition with particles occupying all even
integers.\\[0.5em]
2. Step initial condition with particles occupying all negative
integers.\\[0.5em]
It is not obvious that the infinite-particle PushASEP started from
these initial configurations is correctly defined, and some work
needs to be done to prove the existence of the Markovian dynamics.
However, we take a simpler path here and consider our
infinite-particle system as a limit of growing finite-particle
systems. It turns out that for the above initial conditions, the
distribution of any finite number of particles at any finitely many
time moments stabilizes as the total number of particles in the
system becomes large enough. It is this limiting distribution that
we analyze.

We are able to control the asymptotic behavior of the joint
distribution of $x_{n_1}(t_1), \dots,x_{n_k}(t_k)$ with
$x_{n_1}(0)\ge \dots\ge x_{n_k}(0)$ and $t_1\ge\dots \ge t_k$. It is
the second main novel feature of the present paper (the first one
being the model itself) that we can handle joint distributions of
different particles at different time moments. As special cases we
find distributions of several particles at a given time moment and
distribution of one particle at several time moments (a.k.a. the
tagged particle).

In the growth model formulation of PushASEP (that we do not give
here; it can be easily reconstructed from the growth models for
TASEP and Toom's model described in \cite{DLSS91} and references therein), this
corresponds to joint distributions of values of the height function
at a finite number of space-time points that lie on a space-like
path; for that reason we use the term `space-like path' below. The
two extreme space-like paths were described above -- they correspond
to $t_1=\dots=t_k$ and $n_1=\dots=n_k$.

The algebraic techniques of handling space-like paths are used in the subsequent paper~\cite{BFS07b} to analyze two different models, namely the polynuclear growth (PNG) model on a flat substrate and TASEP in discrete time with parallel update.

Our main result states that large time fluctuations of the particle
positions along any space-like path have exponents $1/3$ and $2/3$,
and that the limiting process is the Airy$_1$ process for the flat
initial condition and the Airy$_2$ process for the step initial
condition (see the review~\cite{Fer07} and Section~\ref{SectLimitProcesses}
below for the definition of these processes).

In the PushASEP model, we have the fluctuation exponent $1/3$ even in the case of zero drift. This is due to the asymmetry in the dynamical rules and it is consistent with the KPZ hypothesis. In fact, from KPZ we expect to have the $1/3$ exponent when $j''(\rho)\neq 0$, where $j(\rho)$ is the current of particles as a function of their density $\rho$, and $j''(\rho)=-2(R+L/(1-\rho)^3)$ for PushASEP.

We find it remarkable that up to scaling factors, the fluctuations
are independent of the space-like path we choose (this phenomenon
was also observed in \cite{BO04} for the polynuclear
growth model (PNG) with step initial condition). It is natural to
conjecture that this type of universality holds at least as broadly
as KPZ-universality does.

Interestingly enough, so far it is unknown how to study the joint
distribution of $x_{n_1}(t_1)$ and $x_{n_2}(t_2)$ with
$x_{n_1}(0)>x_{n_2}(0)$ and $t_1<t_2$ (two points on a time-like
path); this question remains a major open problem of the subject.

\vspace{0.5em}
\emph{Previous results.} For the TASEP and PNG models, large time fluctuation results have already been obtained in the following cases: For the step initial condition the Airy$_2$ process has been shown to occur in the scaling limit for fixed time~\cite{PS02,Jo03b,Jo03}, and more recently for tagged particle~\cite{SI07}. For TASEP, the Airy$_1$ process occurs for flat initial conditions in continuous time~\cite{BFPS06} and in discrete time with sequential update~\cite{BFP06} with generalization to the initial condition of one particle every $d\geq 2$ sites\footnote{Similar results for discrete time TASEP with parallel update and PNG model will follow from
more general results of~\cite{BFS07b}.}. Also, a transition between the Airy$_2$ and Airy$_1$ processes was obtained in~\cite{BFS07}. These are fixed time results; the only previous result concerning general space-like paths is to be found in~\cite{BO04} in the context of the PNG model, where the Airy$_2$ process was obtained as a limit for a directed percolation model.

\vspace{0.5em}
\emph{Outline.} The paper is organized as follows. In Section~\ref{sectModel} we describe the model and the results. In Proposition~\ref{Prop1} the transition probability of the model is given. Then, we define what we mean by space-like paths, and formulate the scaling limit results; the definitions of the Airy$_1$ and Airy$_2$ processes are recalled in Section~\ref{SectLimitProcesses}. In Section~\ref{SectKernels} we state the general kernel for PushASEP (Proposition~\ref{PropPushASEP}) and then particularize it to step and flat initial conditions (Proposition~\ref{PropStepIC} and~\ref{PropFlatIC}). In Section~\ref{SectDetMeasures} we first prove Proposition~\ref{Prop1} and then obtain the general kernel for a determinantal measure of a certain form (Theorem~\ref{ThmKernel}), which includes the one of PushASEP. Finally, the asymptotic analysis is the content of Section~\ref{SectAsympt}.

\subsubsection*{Acknowledgments}
We are very grateful to the anonymous referee for careful reading
and a number of constructive remarks. A.Borodin was partially
supported by the NSF grants DMS-0402047 and DMS-0707163.

\newpage
\section{The PushASEP model and limit results}\label{sectModel}
\subsection{The PushASEP}
The model we consider is an extension of the well known totally asymmetric simple exclusion process (TASEP) on $\Z$. The allowed configuration are like in the TASEP, i.e., configurations consist of particles on $\Z$, with the constraint that at each site can be occupied by at most one particle (exclusion constraint). We consider a dynamics in continuous time, where particles are allowed to jump to the right and to the left as follows. A particle jumps to its right-neighbor site with some rate, provided the site is empty (TASEP dynamics). To the left, a particle jump to its left-neighbor site with some rate and, if the site is already occupied by another particle, this is pushed to its left-neighbor and so on (push dynamics).

To define precisely the jump rates, we need to introduce a few
notations. Since the dynamics preserves the relative order of
particles, we can associate to each particle a label. Let $x_k(t)$
be the position of particle $k$ at time $t$. We choose the
right-left labeling, i.e., $x_k(t)>x_{k+1}(t)$ for all $k\in
I\subseteq\Z$, $t\geq 0$. With this labeling, we consider $v_k>0$,
$k\in I$, and some smooth positive increasing functions $a(t), b(t)$
with $a(0)=b(0)=0$. Then, the right jump rate of particle $k$ is
$\dot{a}(t) v_k$, while its left jump rate is $\dot{b}(t)/v_k$.

In Proposition~\ref{Prop1} we derive the expression of the transition probability from time $t=0$ to time $t$ for $N$ particles, proven in Section~\ref{SectDetMeasures}.
\begin{prop}\label{Prop1}
Consider $N$ particles with initial conditions $x_i(0)=y_i$. Denote its transition probability until time $t$ by
\begin{equation}\label{eqG}
G(x_N,\ldots,x_1;t)=\Pb(x_i(t)=x_i,1\leq i\leq N | x_i(0)=y_i,1\leq i\leq N).
\end{equation}
Then
\begin{eqnarray}\label{eqProp1a}
&&\hspace{-1.5em} G(x_N,\ldots,x_1;t)\\
&&\hspace{-1.5em} =\bigg(\prod_{n=1}^N v_n^{x_n-y_n} e^{-a(t) v_n} e^{-b(t)/v_n}\bigg) \det\left[F_{k,l}(x_{N+1-l}-y_{N+1-k},a(t),b(t))\right]_{1\leq k,l\leq N}, \nonumber
\end{eqnarray}
where
\begin{equation}\label{eqProp1b}
F_{k,l}(x,a,b)=\frac{1}{2\pi \I}\oint_{\Gamma_0} \dx z z^{x-1} \frac{\prod_{i=1}^{k-1}(1-v_{N+1-i}z)}{\prod_{j=1}^{l-1}(1-v_{N+1-j}z)} e^{b z} e^{a/z},
\end{equation}
where $\Gamma_0$ is any anticlockwise oriented simple loop with including only the pole at $z=0$.
\end{prop}

\subsection{Space-like paths}
The computation of the joint distribution of particle positions at a
given time $t$ can be obtained from Proposition~\ref{Prop1} by
adapting the method used in~\cite{BFPS06} for the TASEP. However,
one of the main motivation for this work is to enlarge the spectrum
of the situations which can be analyzed to what we call
\emph{space-like} paths. In this context, space-like paths are
sequences of particle numbers and times in the ensemble
\begin{equation}
{\cal S}=\{(n_k,t_k),k\geq 1 | (n_k,t_k)\prec (n_{k+1},t_{k+1})\},
\end{equation}
where, by definition,
\begin{equation}\label{DefPrec}
(n_i,t_i)\prec (n_j,t_j)\textrm{ if }n_j\geq n_i, t_j\leq t_i, \textrm{ and the two couples are not identical.}
\end{equation}
The two extreme cases are (1) fixed time, $t_k=t$ for all $k$, and (2) fixed particle number, $n_k=n$ for all $k$. This last situation is known as \emph{tagged particle} problem. Since the analysis is of the same degree of difficulty for any space-like path, we will consider the general situation.

\begin{figure}
\begin{center}
\psfrag{n}{$n$}
\psfrag{t}{$t$}
\psfrag{w0}{$w^0$}
\psfrag{w1}{$w^1$}
\psfrag{f}{$\pi$}
\includegraphics[height=5cm]{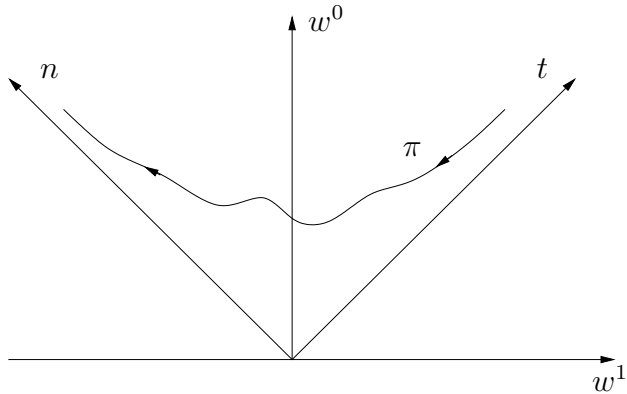}
\caption{An example of a space-like path. Its slope is, in absolute value, at most $1$.}
\label{FigSpaceLikePath}
\end{center}
\end{figure}

Consider any smooth function $\pi$, $w^0=\pi(w^1)$, in the forward light cone of the origin that satisfies
\begin{equation}
|\pi' | \leq 1,\quad |w^1|\leq \pi(w^1).
\end{equation}
These are space-like paths in $\R\times \R_+$, see Figure~\ref{FigSpaceLikePath}.
The first condition (the space-like property) is related to the applicability of our result to sequences of particles in $\cal S$. The second condition just reflects the choice of having $t\geq 0$ and $n\geq 0$. Time and particle number are connected with the variables $w^1$ and $w^0$ by a rotation of $45$ degrees. To avoid unnecessary $\sqrt{2}$'s, we set
\begin{equation}
\left\{\begin{array}{rcl}
w^1&=&\frac{t-n}{2}\\
w^0&=&\frac{t+n}{2}
\end{array}\right\}
\Longleftrightarrow
\left\{\begin{array}{rcl}
t&=&w^0+w^1\\
n&=&w^0-w^1
\end{array}\right\}
\end{equation}

We want to study the joint distributions of particle positions in
the limit of large time, where universal processes arise. Since we
consider several times, we can not simply use $t$ as large
parameter. Instead, we consider a large parameter $T$. Particle
numbers and times under investigation will have a leading term
proportional to $T$. In the $(w^1,w^0)$ plane, we consider $w^1$
around $\theta T$ for a fixed $\theta$, while $w^0=T \pi(w^1/T)$.
From KPZ we know that correlations are on $T^{2/3}$ scale.
Therefore, we set the scaling as
\begin{equation}
\begin{cases}
w^1(u)= \theta T - u T^{2/3},\\
w^0(u)=\pi(\theta) T -\pi'(\theta) u T^{2/3}+\tfrac12 \pi''(\theta) u^2 T^{1/3}.
\end{cases}
\end{equation}
Notice that $w^0(u)$ is equal to $T \pi(w^1(u)/T)$ up to terms that
remain bounded, and they become irrelevant in the large $T$ limit,
since the fluctuations grow as $T^{1/3}$. Coming back to the $(n,t)$
variables, we have
\begin{eqnarray}\label{eqScaling1}
t(u)&=&(\pi(\theta)+\theta) T -(\pi'(\theta)+1) u T^{2/3} + \tfrac12 \pi''(\theta) u^2 T^{1/3},\nonumber \\
n(u)&=&\big[(\pi(\theta)-\theta) T +(1-\pi'(\theta)) u T^{2/3} + \tfrac12 \pi''(\theta) u^2 T^{1/3}\big].
\end{eqnarray}
In particular, setting $\pi(\theta)=1-\theta$ we get the fixed time case with $t=T$, while setting $\pi(\theta)=\alpha+\theta$ we get the tagged particle situation with particle number $n=\alpha T$.

\subsection{Scaling limits}
Universality occurs in the large $T$ limit. In Proposition~\ref{PropPushASEP} we obtain an expression for the joint distribution in the general setting. For the asymptotic analysis we consider the case where all particles have the same jump rates, i.e., we set
\begin{equation}
v_k=1 \textrm{ for all } k\in I.
\end{equation}
Moreover, we consider time-homogeneous case, i.e., we set $a(t)=R t$ and $b(t)=L t$ for some $R,L\geq 0$ (for time non-homogeneous case, one would just replace $R$ and $L$ by some time-dependent functions).
Two important initial conditions are\\[0.5em]
(a) \emph{flat initial condition}: particles start from $2\Z$,\\[0.5em]
(b) \emph{step initial condition}: particles start from $\Z_-=\{\ldots,-3,-2,-1\}$.\\[0.5em]
In the first case, the macroscopic limit shape is flat, while in the second case it is curved, see~\cite{Fer07} for a review on universality in the TASEP. For TASEP with step initial conditions and particle-dependent rates $v_k$, the study of tagged particle has been carried out in~\cite{SI07}.

\subsubsection*{Flat initial conditions}
For the flat initial condition, it is not very difficult to get the proper scaling limit as $T\to\infty$. The initial position of particle $n(u)$ is $-2n(u)$ and during time $t(u)$ it will have traveled around $\mathbf{v}\, t(u)$, where $\mathbf{v}$ is the mean speed of particles, given by
\begin{equation}
\mathbf{v}=-2L+R/2.
\end{equation}
The reason is that the density of particle is $1/2$ and the particles jumps to the right with rate $R$ but the site on its right has a $1/2$ chance to be empty. Moreover, particles move (and push) to the left with rate $L$ but typically every second move to the left is due to a push from another particle. Therefore, the rescaled process is given by
\begin{equation}
u\mapsto X_T(u)=\frac{x_{n(u)}(t(u))-(-2n(u)+\mathbf{v}\, t(u))}{-T^{1/3}},
\end{equation}
where $n(u)$ and $t(u)$ are defined in  (\ref{eqScaling1}). The rescaled process $X_T$ has a limit for large $T$ given in terms of the Airy$_1$ process, $\Af$ (see~\cite{BFPS06,BFS07,Fer07} and Section~\ref{SectLimitProcesses} for details on $\Af$).

\begin{thm}[Convergence to the Airy$_1$ process]\label{ThmFlatCvg}
Let us set the vertical and horizontal rescaling
\begin{equation}\label{eqS}
S_v=((8L+R)(\pi(\theta)+\theta))^{1/3},\quad S_h=\frac{4((8L+R)(\pi(\theta)+\theta))^{2/3}}{(R+4L)(\pi'(\theta)+1)+4(1-\pi'(\theta))}.
\end{equation}
Then
\begin{equation}
\lim_{T\to\infty} X_T(u)=S_v \Af (u/S_h)
\end{equation}
in the sense of finite dimensional distributions.
\end{thm}
The proof of this theorem is in Section~\ref{SectAsympt}. The specialization for fixed time $t=T$ is
\begin{equation}
S_v=(8L+R)^{1/3},\quad S_h=\frac{(8L+R)^{2/3}}{2},
\end{equation}
and the one for tagged particle $n=\alpha T$ at times $t(u)=T-2u T^{2/3}$, obtained by setting $\theta=(1-\alpha)/2$, is
\begin{equation}
S_v=(8L+R)^{1/3},\quad S_h=\frac{2(8L+R)^{2/3}}{4L+R}.
\end{equation}

\subsubsection*{Step initial condition}
\begin{figure}
\begin{center}
\includegraphics[height=5cm]{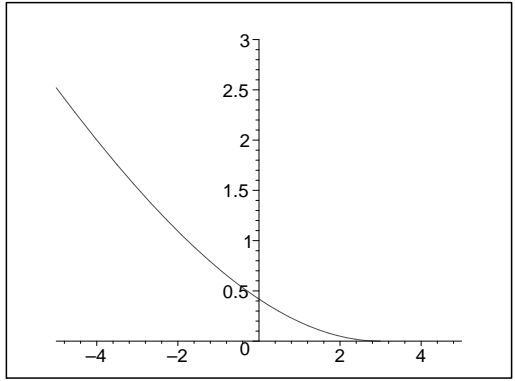}
\caption{Parametric plot of $(\beta(\mu),\alpha(\mu))$, for $L=1, R=4$.}
\label{FigStepIC}
\end{center}
\end{figure}
The proper rescaled process for step initial condition is quite intricate. Denote by $\beta t$ the typical position of particle with number around $\alpha t$ at time $t$. In the situations previously studied in the literature, there was a nice function $\beta=\beta(\alpha)$. In the present situation this is not anymore true, but we can still describe the limit shape. More precisely, $\alpha$ and $\beta$ are parametrized by a $\mu\in (0,1)$ via
\begin{equation}
\alpha(\mu)=(1-\mu)^2(R+L/\mu^2),\quad \beta(\mu)=-((1-2\mu)R+L/\mu^2).
\end{equation}
The parameter $\mu$ comes from the asymptotic analysis in
Section~\ref{SectStepIC}, where it represents the position of the
double critical point. To see that it is a proper parametrization,
we have to verify that for a given point on the space-like curve
$(\theta,\pi(\theta))$ there corresponds exactly one value of $\mu$.
From (\ref{eqScaling1}) we have $n\simeq
t(\pi(\theta)-\theta)/(\pi(\theta)+\theta)$ and, since we have set
$n\simeq \alpha t$, we have
\begin{equation}\label{eq4.6}
\alpha(\mu)=\frac{\pi(\theta)-\theta}{\pi(\theta)+\theta}.
\end{equation}
For any given $\theta$, there exists only one $\mu$ such that (\ref{eq4.6}) holds, because $\alpha$ is strictly monotone in $\mu$. Some computations are needed, but finally we get the rescaling of the position $x$ as a function of $u$, namely,
\begin{equation}\label{eqScaling2}
x(u)=\sigma_0 T -\sigma_1 u T^{2/3} + \sigma_2 u^2 T^{1/3},
\end{equation}
where
\begin{eqnarray}
\sigma_0&=&(\pi(\theta)+\theta)\beta(\mu) \nonumber \\
\sigma_1&=&1+(\pi'(\theta)+1)\left(\mu R-\tfrac{L}{\mu}\right)+(1-\pi'(\theta))\frac{1}{1-\mu} \\
\sigma_2&=&\tfrac12 \pi''(\theta)\left(\mu R+\tfrac{L}{\mu}-\tfrac{1}{1-\mu}\right)+ \frac{(\pi'(\theta)(1-\alpha(\mu))-(1+\alpha(\mu)))^2}{4(1-\mu)^3(\pi(\theta)+\theta)(R+L/\mu^3)}. \nonumber
\end{eqnarray}
The rescaled process is then given by
\begin{equation}
u\mapsto X_T(u)=\frac{x_{n(u)}(t(u))-(\sigma_0 T -\sigma_1 u T^{2/3} + \sigma_2 u^2 T^{1/3})}{-T^{1/3}},
\end{equation}
with $n(u)$ and $t(u)$ given in (\ref{eqScaling1}). Define the constants
\begin{eqnarray}
\kappa_0&=&\frac{(\pi(\theta)+\theta)(R+L/\mu^3)}{\mu(1-\mu)},\nonumber \\
\kappa_1&=&\frac{(\pi'(\theta)+1)(R+L/\mu^2)}{2\mu}-\frac{\pi'(\theta)-1}{2\mu(1-\mu)^2}.
\end{eqnarray}
Then, a detailed asymptotic analysis would lead to,
\begin{equation}\label{eq4.11}
\lim_{T\to\infty}X_T(u)=\mu\kappa_0^{1/3}\Ac(\kappa_1 \kappa_0^{-2/3} u),
\end{equation}
in the sense of finite dimensional distributions, where $\Ac$ is the Airy$_2$ process (see~\cite{PS02,Jo03b,Fer07} and Section~\ref{SectLimitProcesses} for details on $\Ac$). As for the flat PNG, special cases are tagged particle and fixed time. In Section~\ref{SectStepIC} we obtain (\ref{eq4.11}) by looking at the contribution coming from the series expansion around a double critical point. To get (\ref{eq4.11}) rigorously, one has to control (1) the error terms in the convergence on bounded sets and (2) get some bounds to get convergence of the Fredholm determinants. This is what we actually do in the flat initial condition setting.

\subsection{Limit processes}\label{SectLimitProcesses}
For completeness, we shortly recall the definitions of the limit processes $\Af$ and $\Ac$ appearing above. The notation $\Ai(x)$ below stands for the classical Airy function~\cite{AS84}.
\begin{defin}[The Airy$_1$ process]
The \emph{{Airy$_1$ process}} $\Af$ is the process with $m$-point joint distributions at \mbox{$u_1< u_2< \ldots <u_m$} given by the Fredholm determinant
\begin{equation}\label{eqFredDetAiry1}
\Pb\Big(\bigcap_{k=1}^m\{\Af(u_k)\leq s_k\}\Big)=
\det(\Id-\chi_s K_{\Af}\chi_s)_{L^2(\{u_1,\ldots,u_m\}\times\R)},
\end{equation}
where $\chi_s(u_k,x)=\Id(x>s_k)$ and the kernel $K_{\Af}$ is given by
\begin{eqnarray}\label{eqKernelExpanded}
& &\hspace{-3em} K_{\Af}(u_1,s_1;u_2,s_2)=-\frac{1}{\sqrt{4\pi (u_2-u_1)}}\exp\left(-\frac{(s_2-s_1)^2}{4 (u_2-u_1)}\right) \Id(u_2>u_1) \nonumber \\
& &\hspace{-3em} + \Ai(s_1+s_2+(u_2-u_1)^2) \exp\left((u_2-u_1)(s_1+s_2)+\frac23(u_2-u_1)^3\right).
\end{eqnarray}
\end{defin}

\begin{defin}[The Airy$_2$ process]
The \emph{{Airy$_2$ process}} $\Ac$ is the process with $m$-point joint distributions at \mbox{$u_1< u_2< \ldots <u_m$} given by the Fredholm determinant
\begin{equation}\label{eqFredDetAiry2}
\Pb\Big(\bigcap_{k=1}^m\{\Ac(u_k)\leq s_k\}\Big)=
\det(\Id-\chi_s K_{\Af}\chi_s)_{L^2(\{u_1,\ldots,u_m\}\times\R)},
\end{equation}
where $\chi_s(u_k,x)=\Id(x>s_k)$ and the kernel $K_{\Ac}$ is given by
\begin{equation}\label{eqKernelExpanded2}
K_{\Ac}(u_1,s_1;u_2,s_2)=
\begin{cases}
\int_{\R_+} e^{-\lambda (u_2-u_1)}\Ai(s_1+\lambda)\Ai(s_2+\lambda),& u_2\geq u_1,\\
-\int_{\R_-} e^{-\lambda(u_2-u_1)}\Ai(s_1+\lambda)\Ai(s_2+\lambda),& u_2<u_1.
\end{cases}
\end{equation}
\end{defin}

\section{Finite time kernel}\label{SectKernels}
In this section we first derive an expression for the joint
distributions of particle positions in a finite system. They are
given by Fredholm determinants of a kernel, which is first stated
for general jump rates and initial positions. After that, we
specialize to the cases of uniform jump rates in the case of step
and flat initial conditions. Flat initial conditions are obtained
via a limit of finite systems.

\subsection{General kernel for PushASEP}
To state the following result, proven in Section~\ref{SectDetMeasures}, we introduce a space of
functions $V_n$. Consider the set of numbers $\{v_1,\ldots,v_n\}$
and let \mbox{$\{u_1<u_2<\ldots<u_{\nu}\}$} be their different
values, with $\alpha_k$ being the multiplicity of $u_k$ ($v_k$ is
the jump rate of particle with label $k$). Then we define the space
\begin{equation}\label{eqVn}
V_n=\mathrm{span}\{x^l u_k^x ,1\leq k \leq \nu, 0\leq l \leq \alpha_k-1\}.
\end{equation}
Recall that the evolution of particle indexed by $n$ is independent
of the particles with index $m>n$.
\begin{prop}\label{PropPushASEP}
Consider a system of particles with indices $n=1,2,\ldots$ starting
from positions $y_1>y_2>\ldots$. Denote by $x_n(t)$ the position of
particle with index $n$ at time $t$. Then the joint distribution of
particle positions is given by the Fredholm determinant
\begin{equation}\label{eqOld3.1}
\Pb\Big(\bigcap_{k=1}^m \{x_{n_k}(t_k)\geq s_k\}\Big)=
\det\big(\Id-\tilde\chi_s K \tilde\chi_s\big)_{\ell^2(\{(n_1,t_1),\ldots,(n_m,t_m)\}\times\Z)}
\end{equation}
with $((n_1,t_1),\ldots,(n_m,t_m))\in {\cal S}$, and
$\tilde\chi_s((n_k,t_k))(x)=\Id(x<s_k)$. The kernel $K$ is given by
\begin{equation}\label{3.4}
K((n_1,t_1),x_1; (n_2,t_2),x_2)=-\phi^{((n_1,t_1),(n_2,t_2))}(x_1,x_2)+
\sum_{k=1}^{n_2} \Psi^{n_1,t_1}_{n_1-k}(x_1) \Phi^{n_2,t_2}_{n_2-k}(x_2)
\end{equation}
where
\begin{equation}\label{eqPSI}
\Psi^{n,t}_{n-l}(x) = \frac{1}{2\pi\I}\oint_{\Gamma_0}\dx z
z^{x-y_l-1} e^{a(t)/z+b(t)z}\frac{(1-v_1 z)\cdots(1-v_n z)}{(1-v_1
z)\cdots (1-v_l z)}, \quad l=1,2,\dots,
\end{equation}
the functions ${\{\Phi^{n,t}_{n-k}\}}_{k=1}^n$ are uniquely determined by the orthogonality relations
\begin{equation}\label{eqOrthoCond}
\sum_{x\in\Z}\Psi^{n,t}_{n-l}(x)\Phi^{n,t}_{n-k}(x)=\delta_{k,l}, \quad 1\leq k,l\leq n,
\end{equation}
and by the requirement
$\mathrm{span}\{\Phi^{n,t}_{n-l}(x),l=1,\ldots,n\} = V_n$.
The first term in (\ref{3.4}) is given by
\begin{equation}\label{eqTrans}
\phi^{((n_1,t_1),(n_2,t_2))}(x,y)=
\frac{1}{2\pi\I}\oint_{\Gamma_0}\frac{\dx z}{z^{y-x+1}} \frac{e^{(a(t_1)-a(t_2))/z}e^{(b(t_1)-b(t_2))z}}{(1-v_{n_1+1}z)\cdots (1-v_{n_2}z)} \Id_{[(n_1,t_1)\prec (n_2,t_2)]}.
\end{equation}
The notation $\Gamma_0$ stands for any anticlockwise oriented simple loop including only the pole at $0$.
\end{prop}

\subsection{Kernel for step initial condition}
We set all the jump rates to $1$: $v_1=v_2=\cdots=1$. The transition function (\ref{eqTrans}) does not depend on initial conditions. It is useful to rewrite it in a slightly different form.
\begin{lem}\label{lemTransFct}
The transition function can be rewritten as
\begin{eqnarray}\label{eqTransFct}
& & \phi^{((n_1,t_1),(n_2,t_2))}(x,y) \\
&=&\frac{1}{2\pi\I}\oint_{\Gamma_{0,1}}\dx w \frac{1}{w^{x-y+1}} \left(\frac{w}{w-1}\right)^{n_2-n_1} \frac{e^{a(t_1)w+b(t_1)/w}}{e^{a(t_2)w+b(t_2)/w}} \Id_{[(n_1,t_1)\prec (n_2,t_2)]}.\nonumber
\end{eqnarray}
\end{lem}
\begin{proofOF}{Lemma~\ref{lemTransFct}}
The proof follows by the change of variable $z=1/w$ in (\ref{eqTrans}).
\end{proofOF}

\begin{lem}\label{LemOrthoStepIC}
Let $y_i=-i,i\geq 1$. Then, the functions $\Phi$ and $\Psi$ are given by
\begin{eqnarray}
\Psi^{n,t}_{k}(x)&=& \frac{1}{2\pi\I}\oint_{\Gamma_{0,1}}\dx w \frac{(w-1)^k}{w^{x+n+1}}e^{a(t)w + b(t)/w},\nonumber\\
\Phi^{n,t}_{j}(x)&=& \frac{1}{2\pi\I}\oint_{\Gamma_{1}}\dx z \frac{z^{x+n}}{(z-1)^{j+1}}e^{-a(t)z - b(t)/z}.
\end{eqnarray}
\end{lem}
\begin{proofOF}{Lemma~\ref{LemOrthoStepIC}}
$\Psi^{n,t}_k(x)$ comes from the change of variable $z=1/w$ in
(\ref{eqPSI}). For $k\geq 0$, the pole at $w=1$ is irrelevant, but
in the kernel $\Psi_k^{n,t}$ enters also for negative values of $k$.

We have to verify that the function $\Phi^{n,t}_j(x)$ satisfy the orthogonal condition (\ref{eqOrthoCond}) and span the space $V_n$ given in (\ref{eqVn}). For $v_1=\cdots=v_n=1$, $V_n={\rm span}(1,x,\ldots,x^{n-1})$. By the residue's theorem, the function $\Phi^{n,t}_j(x)$ is a polynomial of degree $j$ in $x$.
Thus, ${\rm span}(\Phi^{n,t}_j(x),j=0,\ldots,n-1)=V_n$.

The second step is to compute $\sum_{x\in\Z}\Phi^{n,t}_j(x)\Psi^{n,t}_k(x)$ for $0\leq j,k\leq n-1$.
We divide it into the sum over $x\geq 0$ and the one over $x<0$. We have
\begin{equation}\label{eq3.9}
\sum_{x\geq 0}\Phi^{n,t}_j(x)\Psi^{n,t}_k(x)=
\sum_{x\geq 0} \frac{1}{(2\pi\I)^2}\oint_{\Gamma_1}\dx z \oint_{\Gamma_0}\dx w \frac{e^{a(t)w+b(t)/w}}{e^{a(t)z+b(t)/z}}\frac{(w-1)^k}{(z-1)^{j+1}} \frac{z^{x+n}}{w^{x+n+1}}.
\end{equation}
We choose the paths $\Gamma_0$ and $\Gamma_1$ satisfying $|z|<|w|$, so that we can take the sum inside the integrals and use
\begin{equation}
\sum_{x\geq 0} \frac{z^x}{w^{x+1}}=\frac{1}{w-z},
\end{equation}
to get
\begin{equation}\label{eq3.11}
(\ref{eq3.9})= \frac{1}{(2\pi\I)^2}\oint_{\Gamma_1}\dx z \oint_{\Gamma_{0,z}}\dx w \frac{e^{a(t)w+b(t)/w}}{e^{a(t)z+b(t)/z}}\frac{(w-1)^k}{(z-1)^{j+1}} \frac{z^{n}}{w^{n}}\frac{1}{w-z},
\end{equation}
where the subscript $z$ in $\Gamma_{0,z}$ reminds that $z$ is a pole for the integral over $w$.

Next consider the sum over $x<0$. We have
\begin{equation}\label{eq3.9b}
\sum_{x< 0}\Phi^{n,t}_j(x)\Psi^{n,t}_k(x)=
\sum_{x< 0} \frac{1}{(2\pi\I)^2} \oint_{\Gamma_0}\dx w \oint_{\Gamma_1}\dx z \frac{e^{a(t)w+b(t)/w}}{e^{a(t)z+b(t)/z}}\frac{(w-1)^k}{(z-1)^{j+1}} \frac{z^{x+n}}{w^{x+n+1}}.
\end{equation}
This time we choose the paths $\Gamma_0$ and $\Gamma_1$ satisfying $|z|>|w|$ and then take the sum inside the integrals. Using
\begin{equation}
\sum_{x<0} \frac{z^x}{w^{x+1}}=-\frac{1}{w-z}
\end{equation}
we obtain
\begin{equation}\label{eq3.11b}
(\ref{eq3.9b})= - \frac{1}{(2\pi\I)^2} \oint_{\Gamma_{0}}\dx w \oint_{\Gamma_{1,w}}\dx z \frac{e^{a(t)w+b(t)/w}}{e^{a(t)z+b(t)/z}}\frac{(w-1)^k}{(z-1)^{j+1}} \frac{z^{n}}{w^{n}}\frac{1}{w-z},
\end{equation}
where now $w$ is a pole for the integral over $z$. Thus,
\begin{equation}
\sum_{x\in\Z}\Phi^{n,t}_j(x)\Psi^{n,t}_k(x) = (\ref{eq3.11})+(\ref{eq3.11b}).
\end{equation}
We can deform the paths of integration in (\ref{eq3.11b}) so that
they become as the integration paths of (\ref{eq3.11}) up to
correcting the contribution of the residue at $z=w$. Thus we finally
get, for $0\leq j,k\leq n-1$,
\begin{equation}
\sum_{x\in\Z}\Phi^{n,t}_j(x)\Psi^{n,t}_k(x)= \frac{1}{2\pi\I}\oint_{\Gamma_1}\dx z (z-1)^{k-j-1}=\delta_{j,k}.
\end{equation}
As a side remark, the same computations would not hold for $k,j<0$, for which $\Phi^{n,t}_j(x)\equiv 0$, because it is not possible to choose the paths with $|z|>|w|$ without introducing an extra pole at $z=0$.
\end{proofOF}

\begin{prop}[Step initial condition, finite time kernel]\label{PropStepIC}
$ $ \\The kernel for $y_i=-i$, $i\geq 1$, is given by
\begin{eqnarray}\label{eqKernelStep}
& & K((n_1,t_1),x_1; (n_2,t_2),x_2) \\
&=&-\frac{1}{2\pi\I}\oint_{\Gamma_{0}}\dx w \frac{1}{w^{x_1-x_2+1}} \left(\frac{w}{1-w}\right)^{n_2-n_1} \frac{e^{a(t_1)w+b(t_1)/w}}{e^{a(t_2)w+b(t_2)/w}} \Id_{[(n_1,t_1)\prec (n_2,t_2)]} \nonumber \\
&+&\frac{1}{(2\pi\I)^2}\oint_{\Gamma_0}\dx w \oint_{\Gamma_1}\dx z \frac{e^{b(t_1)/w+a(t_1)w}}{e^{b(t_2)/z+a(t_2)z}} \frac{(1-w)^{n_1}}{w^{x_1+n_1+1}} \frac{z^{x_2+n_2}}{(1-z)^{n_2}}\frac{1}{w-z}. \nonumber
\end{eqnarray}
The contours $\Gamma_0$ and $\Gamma_1$ include the poles $w=0$ and $z=1$ and no other poles. This means in particular that $\Gamma_0$ and $\Gamma_1$ are disjoints, because of the term $1/(w-z)$.
\end{prop}
\begin{proofOF}{Proposition~\ref{PropStepIC}}
Consider the main term of the kernel, namely
\begin{multline}\label{eq3.18}
\sum_{k=1}^{n_2}\Psi^{n_1,t_1}_{n_1-k}(x_1)\Phi^{n_2,t_2}_{n_2-k}(x_2)
=\sum_{k=1}^{n_2} \frac{1}{2\pi\I}\oint_{\Gamma_{0,1}}\dx w \frac{(w-1)^{n_1-k}}{w^{x_1+n_1+1}}e^{a(t_1)w + b(t_1)/w}\\
\times \frac{1}{2\pi\I}\oint_{\Gamma_{1}}\dx z \frac{z^{x_2+n_2}}{(z-1)^{n_2-k+1}}e^{-a(t_2)z - b(t_2)/z}.
\end{multline}
First we extend the sum to $+\infty$, since the second term is identically equal to zero for $k>n_2$. We choose the integration paths so that $|z-1|<|w-1|$. Then, we can take the sum inside the integral. The $k$-dependent terms are
\begin{equation}
\sum_{k\geq 1} \frac{(z-1)^{k-1}}{(w-1)^{k}}=\frac{1}{w-z}.
\end{equation}
Thus, we get
\begin{multline}\label{eq3.20}
(\ref{eq3.18})=\frac{1}{(2\pi\I)^2} \oint_{\Gamma_{1}}\dx z \oint_{\Gamma_{0,z}}\dx w\frac{e^{a(t_1)w + b(t_1)/w}}{e^{a(t_2)z+b(t_2)/z}} \frac{(w-1)^{n_1}}{w^{x_1+n_1+1}} \frac{z^{x_2+n_2}}{(z-1)^{n_2}}\frac{1}{w-z}.
\end{multline}

Notice now we have a new pole at $w=z$, but the one at $w=1$ vanished. The contribution of the pole at $w=z$ is exactly equal to the contribution of the pole at $z=1$ in the transition function (\ref{eqTransFct}). Therefore in the final result the first term coming from (\ref{eqTransFct}) has the integral only around $z=0$, and the second term is (\ref{eq3.20}) but with the integral over $w$ only around the pole at $w=0$ and does not contain $z$. Finally, a conjugation by a factor $(-1)^{n_1-n_2}$ gives the final result.
\end{proofOF}

\subsection{Kernel for flat initial condition}
The kernel for the flat initial condition is obtained as a limit of those for systems with finitely many particles as follows. We first compute the kernel for a finite number of particles starting from $y_i=-2i$, $i\geq 1$. Then we shift the focus by $N$ particles, i.e., we consider particles with numbers $N+n_i$ instead of those with numbers $n_i$. For any finite time $t$, we then take the $N\to\infty$ limit, in which the deviations due to the finite number of particles on the right tend to zero. The limiting kernel is what we call the kernel for the flat initial condition ($y_i=-2i$ with $i\in\Z$).

For this case we also consider the homogeneous jump rates, $v_1=v_2=\cdots=1$.
\begin{lem}\label{LemOrthoFlatIC}
Let $y_i=-2i$, $i\geq 1$. Then, the functions $\Phi$ and $\Psi$ are given by
\begin{eqnarray}
\Psi^{n,t}_{k}(x)&=& \frac{1}{2\pi\I}\oint_{\Gamma_{0,1}}\dx w \frac{(w(w-1))^k}{w^{x+2n+1}}e^{a(t)w + b(t)/w},\nonumber\\
\Phi^{n,t}_{j}(x)&=& \frac{1}{2\pi\I}\oint_{\Gamma_{1}}\dx z \frac{(2z-1)z^{x+2n}}{(z(z-1))^{j+1}}e^{-a(t)z - b(t)/z}.
\end{eqnarray}
\end{lem}
\begin{proofOF}{Lemma~\ref{LemOrthoFlatIC}}
The proof is almost identical to the one of Lemma~\ref{LemOrthoStepIC}. The only difference is that contribution of the residue is in this case given by
\begin{eqnarray}
\sum_{x\in\Z}\Phi^{n,t}_j(x)\Psi^{n,t}_k(x)= \frac{1}{2\pi\I}\oint_{\Gamma_1}\dx z (2z-1)(z(z-1))^{k-j-1}=\delta_{j,k}
\end{eqnarray}
by the change of variable $w=z(z-1)$.
\end{proofOF}

\begin{prop}[Flat initial conditions, finite time kernel]\label{PropFlatIC}
$ $ \\
The kernel for $y_i=-2i$, $i\in\Z$, is given by
\begin{eqnarray}\label{eqKernelFlat}
& & K((n_1,t_1),x_1; (n_2,t_2),x_2)\nonumber \\
&=&-\frac{1}{2\pi\I}\oint_{\Gamma_{0}}\dx w \frac{1}{w^{x_1-x_2+1}} \left(\frac{w}{1-w}\right)^{n_2-n_1} \frac{e^{a(t_1)w+b(t_1)/w}}{e^{a(t_2)w+b(t_2)/w}} \Id_{[(n_1,t_1)\prec (n_2,t_2)]} \nonumber\\
&+&\frac{-1}{2\pi\I}\oint_{\Gamma_1}\dx z \frac{e^{a(t_1)(1-z)+b(t_1)/(1-z)}}{e^{a(t_2)z+b(t_2)/z}} \frac{z^{n_1+n_2+x_2}}{(1-z)^{n_1+n_2+x_1+1}}.
\end{eqnarray}
\end{prop}
\begin{proofOF}{Proposition~\ref{PropFlatIC}}
The first step is to get the kernel for $y_i=-2i$, $i\geq 1$. This step is similar to the one of Proposition~\ref{PropStepIC}. The integration paths are taken to satisfy $|z(z-1)|<|w(w-1)|$. Then this time, the sum in $k$ is
\begin{equation}
\sum_{k\geq 1} \frac{(z(z-1))^{k-1}}{(w(w-1))^k} = \frac{1}{(w-z)(w-1+z)}
\end{equation}
and we get
\begin{equation}
\frac{1}{(2\pi\I)^2}\oint_{\Gamma_1}\dx z \oint_{\Gamma_{0,1-z,z}}\hspace{-1em}\dx w \frac{e^{a(t_1)w+b(t_1)/w}}{e^{a(t_2)z+b(t_2)/z}}\frac{(w-1)^{n_1}}{(z-1)^{n_2}}\frac{z^{x_2+n_2}}{w^{x_1+n_1+1}}
\frac{2z-1}{(w-z)(w-1+z)}.
\end{equation}
Notice that the pole for $w=1$ is now replaced by two simple poles, one at $w=z$ and one at $w=1-z$. The pole at $w=z$ cancels with the one at $z=1$ of (\ref{eqTransFct}). Therefore, the kernel for $y_i=-2i$, $i\geq 1$, is given by
\begin{multline}\label{eq3.26}
-\frac{1}{2\pi\I}\oint_{\Gamma_{0}}\dx w \frac{1}{w^{x_1-x_2+1}} \left(\frac{w}{1-w}\right)^{n_2-n_1} \frac{e^{a(t_1)w+b(t_1)/w}}{e^{a(t_2)w+b(t_2)/w}} \Id_{[(n_1,t_1)\prec (n_2,t_2)]}\\
+\frac{1}{(2\pi\I)^2}\oint_{\Gamma_1}\dx z \oint_{\Gamma_{0,1-z}}\hspace{-1em}\dx w \frac{e^{a(t_1)w+b(t_1)/w}}{e^{a(t_2)z+b(t_2)/z}}\frac{(w-1)^{n_1}}{(z-1)^{n_2}}\frac{z^{x_2+n_2}}{w^{x_1+n_1+1}}
\frac{2z-1}{(w-z)(w-1+z)}.
\end{multline}

At this point, we pick a large $N$ and shift the focus to particles
around the $N$th one. Accordingly, we shift the positions by $-2N$.
More precisely, in (\ref{eq3.26}), we replace
\begin{equation}
n_i\to n_i+N,\quad x_i \to x_i-2N.
\end{equation}
Then we get the kernel $K=K_0+K_1+K^{(N)}$ with $((n_1,t_1),x_1; (n_2,t_2),x_2)$-entries given by
\begin{eqnarray}\label{eq3.26b}
K_0&=&-\frac{1}{2\pi\I}\oint_{\Gamma_{0}}\dx w \frac{1}{w^{x_1-x_2+1}} \left(\frac{w}{1-w}\right)^{n_2-n_1} \frac{e^{a(t_1)w+b(t_1)/w}}{e^{a(t_2)w+b(t_2)/w}} \Id_{[(n_1,t_1)\prec (n_2,t_2)]} \nonumber \\
K_1&=&\frac{(-1)^{n_1-n_2+1}}{2\pi\I}\oint_{\Gamma_1}\dx z \frac{e^{a(t_1)(1-z)+b(t_1)/(1-z)}}{e^{a(t_2)z+b(t_2)/z}}\frac{z^{x_2+n_2+n_1}}{(1-z)^{x_1+n_1+n_2+1}},\\
K^{(N)}&=& \frac{1}{(2\pi\I)^2}\oint_{\Gamma_1}\dx z \oint_{\Gamma_0}\dx w \frac{e^{a(t_1)w+b(t_1)/w}}{e^{a(t_2)z+b(t_2)/z}}\frac{(w-1)^{n_1+N}}{(z-1)^{n_2+N}}\frac{z^{x_2+n_2-N}}{w^{x_1+n_1-N+1}}
\frac{2z-1}{(w-z)(w-1+z)}.\nonumber
\end{eqnarray}
The terms $K_0$ and $K_1$ are independent of $N$, while $K^{(N)}$ is not. We need to show that in the $N\to\infty$ limit, the contribution of $K^{(N)}$ vanishes, in the sense that the Fredholm determinant giving the joint distributions of Proposition~\ref{PropPushASEP} converges to the one with kernel $K_0+K_1$.

The Fredholm determinant (\ref{eqOld3.1}) is projected onto $x_i<s_i$. Therefore, for any given $s_1,\ldots,s_m$ we need to get bounds on the kernel for $x_i$'s bounded from above, say for $x_i\leq \ell$ for an $\ell\in\Z$ fixed. In the simplest case of pure TASEP dynamics ($b(t)\equiv 0$), the limit turns out to be easy because for $x_1+n_1 < N$, the pole at $w=0$ vanishes. However, in our model, $b(t)$ is generically non-zero and the integrand has an essential singularity at $w=0$. In what follows, we choose the indices $k\in\{1,\ldots,m\}$ so that $(n_k,t_k)\prec (n_{k+1},t_{k+1})$, $k=1,\ldots,m-1$. Also, we simplify the notation by writing $k\in\{1,\ldots,m\}$ instead of $(n_k,t_k)$ in the arguments of the kernel. Then, the Fredholm determinant becomes
\begin{equation}\label{eq3.29}
(\ref{eqOld3.1})=\sum_{n\geq 0} \frac{(-1)^n}{n!}\sum_{i_1,\ldots,i_n=1}^m \sum_{x_1< s_{i_1}}\cdots  \sum_{x_n< s_{i_n}} \det\left(K(i_k,x_k;i_l,x_l)\right)_{1\leq k,l\leq n}.
\end{equation}
We apply the following conjugation of the kernel, which keeps unchanged the above expression,
\begin{equation}
\widetilde K(i_k,x_k;i_l,x_l) = K(i_k,x_k;i_l,x_l)e^{\e i_l x_l-\e i_k x_k} e^{(x_l-x_k)/2}.
\end{equation}
Using the bound of Lemma~\ref{LemmaBoundsFlat}, for any choice of $\e$ in $(0,(8m)^{-1}]$ and for $x_k,x_l$ bounded from above, we have
\begin{eqnarray}\label{eq3.31}
|\widetilde K_0(i_k,x_k;i_l,x_l)|&\leq& \cte e^{\e x_l},\nonumber \\
|\widetilde K_1(i_k,x_k;i_l,x_l)|&\leq& \cte e^{(x_l+x_k)/8}\leq \cte e^{\e x_l}, \\
|\widetilde K^{(N)}(i_k,x_k;i_l,x_l)|&\leq& \cte e^{(x_l+x_k)/8} \kappa^N\leq \cte e^{\e x_l} \kappa^N, \nonumber
\end{eqnarray}
where $\kappa\in [0,1)$.
In the above bounds, we use the same symbol `$\cte$' for all the constants. With the choice of ordering of the $(n_k,t_k)$'s, we have that $K_0=0=\widetilde K_0$ if $i_l\leq i_k$, thus the bound holds trivially. For the case $i_l>i_k$, Lemma~\ref{LemmaBoundsFlat}  implies the estimate
\begin{equation}
|\widetilde K_0|\leq \cte e^{\e i_k (x_l-x_k)} e^{-|x_l-x_k|/4} e^{\e x_l},
\end{equation}
for $x_k,x_l$ bounded from above. The bound in (\ref{eq3.31}) is then obtained by choosing $\e\leq (4m)^{-1}$, since then $\e i_k\leq \e m \leq 1/4$. The other bounds on $\widetilde K_1$ and $\widetilde K^{(N)}$ are satisfied for $\e m\leq 1/8$.

Therefore, the summand in the multiple sums of (\ref{eq3.29}) is uniformly bounded by
\begin{equation}
\left|\frac{(-1)^n}{n!}\det\left(\widetilde K(i_k,x_k;i_l,x_l)\right)_{1\leq k,l\leq n}\right|\leq \frac{1}{n!}e^{\e(x_1+\ldots+x_n)}\cte^n (1+\kappa^N)^n n^{n/2},
\end{equation}
the term $n^{n/2}$ being Hadamard bound on the value of a $n\times n$ determinant whose entries have modulus bounded by $1$. Since $\kappa<1$, replacing \mbox{$1+\kappa^N$} by $2$ yields a uniform bound, which is summable. Thus, by dominated convergence we can take the $N\to\infty$ limit inside the Fredholm series. Since $\kappa<1$, we have $\lim_{N\to\infty}K^{(N)}=0$, thus the result is proven. Finally, just for convenience, we conjugate the kernel by $(-1)^{n_1-n_2}$, which however has no impact on the Fredholm determinant in question.
\end{proofOF}

\begin{lem}\label{LemmaBoundsFlat}
Let $K_0$, $K_1$, $K^{(N)}$ be as in (\ref{eq3.26b}). Then, for $x_1,x_2\leq \ell$, we have the following bounds.
\begin{eqnarray}
|K_0((n_1,t_1),x_1; (n_2,t_2),x_2)|&\leq& \cte e^{(x_1-x_2)/2} e^{-|x_2-x_1|/4}\Id_{[(n_1,t_1)\prec (n_2,t_2)]}, \nonumber \\
|K_1((n_1,t_1),x_1; (n_2,t_2),x_2)|&\leq& \cte e^{(x_1-x_2)/2} e^{(x_1+x_2)/4},\\
|K^{(N)}((n_1,t_1),x_1; (n_2,t_2),x_2)|&\leq& \cte e^{(x_1-x_2)/2} e^{(x_1+x_2)/4}\kappa^N, \nonumber
\end{eqnarray}
for some $\kappa\in [0,1)$. The constants $\cte$ and $\kappa$ are uniform in $N$ and depend only on $\ell$ and $n_i,t_i$'s.
\end{lem}
\begin{proofOF}{Lemma~\ref{LemmaBoundsFlat}}
For $K_0$ and $x_2-x_1\geq 0$, we can just choose the integration path as $\Gamma_0=\{|w|=e^{-1}\}$, from which we have $|K_0|\leq \cte e^{-(x_2-x_1)}\leq \cte e^{-(x_2-x_1)3/4}$. In the case $x_1-x_2\geq 0$, we choose the integration path as $\Gamma_0=\{|w|=e^{-1/4}\}$. Then, $|K_0|\leq \cte e^{(x_1-x_2)/4}$.

For $K_1$, we choose $\Gamma_1=\{|1-z|=e^{-2}\}$. Then,
\begin{equation}
|K_1|\leq \cte \frac{\max_{\Gamma_1}|z|^{x_2}}{\min_{\Gamma_1} |1-z|^{x_1}}.
\end{equation}
Along $\Gamma_1$, $|1-z|$ is constant, thus $(\min_{\Gamma_1} |1-z|^{x_1})^{-1}=e^{2x_1}\leq \cte e^{3 x_1/4}$ for $x_1$ bounded from above. Remark that $\cte$ depends on the upper bound, $\ell$, for $x_1$. In this case, we can take $\cte=e^{5 \ell /4}$.  Also, for $x_2$ bounded from above, $\max_{\Gamma_1}|z|^{x_2}\leq \cte (1-1/e^2)^{x_2}\leq \cte e^{-x_2/4}$.

For $K^{(N)}$, we use the path $\Gamma_0=\{|w|=e^{-4}\}$ and $\Gamma_1=\{|1-z|=e^{-2}\}$. As required, these paths do not intersect because $1/e^4<1-1/e^2$. Then,
\begin{equation}
|K^{(N)}|\leq \cte \frac{\max_{\Gamma_1}|z|^{x_2}}{\min_{\Gamma_0} |w|^{x_1}} \kappa^N,\quad \kappa=\frac{\max_{\Gamma_0}|w(w-1)|}{\min_{\Gamma_1}|z(z-1)|}.
\end{equation}
For $x_1,x_2$ bounded from above, we have $\max_{\Gamma_1}|z|^{x_2}\leq \cte e^{-x_2/4}$ (as above), and $(\min_{\Gamma_0} |w|^{x_1})^{-1}=e^{4x_1}\leq \cte e^{3x_1/4}$. Finally, it is not difficult to obtain $\kappa=(1+1/e^4)/(1-e^2)=0.159\ldots$, since the maximum of $|w(w-1)|$ is obtained at $w=-e^{-4}$ and the minimum of $|z(z-1)|$ at $z=1-e^{-2}$.
\end{proofOF}

\section{Determinantal measures}\label{SectDetMeasures}
In this section we first prove Proposition~\ref{Prop1}. Then, we use it to extend the measure to space-like paths. More precisely, we first obtain a general determinantal formula in Theorem~\ref{ThmDetMeasure}. Then, in Theorem~\ref{ThmKernel}, we prove that the measure has determinantal correlations and obtain an expression of the associated kernel.

\begin{proofOF}{Proposition~\ref{Prop1}}
We first prove that the initial condition is satisfied. We have
\begin{equation}
F_{k,l}(x,0)=\frac{1}{2\pi \I}\oint_{\Gamma_0} \dx z z^{x-1} \frac{\prod_{i=1}^{k-1}(1-v_{N+1-i}z)}{\prod_{j=1}^{l-1}(1-v_{N+1-j}z)}.
\end{equation}
(a) $F_{k,l}(x,0)=0$ for $x\geq 1$ because the pole at $z=0$ vanishes.\\
(b) $F_{k,l}(x,0)=0$ for $k\geq l$ and $x<l-k$, because then
\begin{equation}
F_{k,l}(x,0)=\frac{1}{2\pi \I}\oint_{\Gamma_0}\dx z z^{x-1} (1-v_l z)\cdots (1-v_{k-1} z)
\end{equation}
and the residue at infinity equals to zero for $x<l-k$.

Assume that $x_N<\cdots<x_1$. If $x_N>y_N$, also $x_l>y_N$ for $l=1,\ldots,N-1$. Thus $F_{1,l}(x_{N+1-l}-y_N,0)=0$ using (a). Therefore $G(x_N,\ldots,x_1;0)=0$. On the other hand, if $x_N<y_N$, then $x_N<y_k-N+k$, $k=1,\ldots,N-1$. Thus $F_{k,1}(x_N-y_{N+1-k},0)=0$ using (b) and the fact that $x_N-y_{N+1-k}<1-k$. Therefore we conclude that $G(x_N,\ldots,x_1;0)=0$ if $x_N\neq y_N$. For $x_N=y_N$, $F_{1,1}(0,0)=1$ and by (a) $F_{1,l}(x_{N+1-l}-y_N,0)=0$ for $l=2,\ldots,N$. This means that
\begin{equation}
G(x_N,\ldots,x_1;0)=\delta_{x_N,y_N} G(x_{N-1},\ldots,x_1;0).
\end{equation}
By iterating the procedure we obtain
\begin{equation}
G(x_N,\ldots,x_1;0)=\prod_{k=1}^N \delta_{x_k,y_k}.
\end{equation}
Notice that the prefactor in (\ref{eqProp1a}) is equal to one at $t=0$.

The initial condition being settled, we need to prove that (\ref{eqProp1a}) satisfies the PushASEP dynamics. For that purpose, let us first compute $\Dt{F_{k,l}(x,t)}{t}$.
\begin{equation}
\Dt{F_{k,l}(x,t)}{t} = \dot{a}(t) F_{k,l}(x-1,t)+\dot{b}(t) F_{k,l}(x+1,t),
\end{equation}
from which it follows, by differentiating the prefactor and the determinant column by column,
\begin{eqnarray}\label{eqDyn1}
\Dt{G(x_N,\ldots,x_1;t)}{t}&=& -\Big(\dot{a}(t) \sum_{k=1}^N v_k + \dot{b}(t) \sum_{k=1}^N \frac{1}{v_k}\Big) G(x_N,\ldots,x_1;t)\nonumber \\
& &+ \dot{a}(t) \sum_{k=1}^N v_k G(\ldots,x_k-1,\ldots;t)\\
& &+\dot{b}(t) \sum_{l=1}^N \frac{1}{v_l} G(\ldots,x_l+1,\ldots;t). \nonumber
\end{eqnarray}
To proceed, we need an identity. Using
\begin{equation}
\frac{z^x}{1-v_{N+1-l}z}=\frac{v_{N+1-l} z^{x+1}}{1-v_{N+1-l} z}+z^x
\end{equation}
it follows that
\begin{equation}\label{eqDyn2}
F_{k,l+1}(x,t)=F_{k,l}(x,t)+v_{N+1-l} F_{k,l+1}(x+1,t).
\end{equation}
Therefore, for $j=2,\ldots,N$, by setting $\tilde y_k=y_{N+1-k}$,
\begin{multline}\label{eq4.7}
G(\ldots,x_j,x_{j-1}=x_j,\ldots;t)
=\frac{1}{Z_N} \det\Big[v_{N+1-l}^{x_{N+1-l}} F_{k,l}(x_{N+1-l}-\tilde y_k,t)\Big]_{1\leq k,l\leq N}\\
= \frac{1}{Z_N} \det\Big[\dots\quad v_j^{x_j} F_{k,N+1-j}(x_j-\tilde y_k,t) \quad
v_{j-1}^{x_j} F_{k,N+2-j}(x_{j-1}-\tilde y_k,t)\cdots \Big].
\end{multline}
Here $Z_N$ does not depend on the $x_j$'s.

Using (\ref{eqDyn2}) we have
\begin{eqnarray}
& & v_{j-1}^{x_j} F_{k,N+2-j}(x_j-\tilde y_k,t) \\
&=& v_{j-1}^{x_j} F_{k,N+1-j}(x_j-\tilde y_k,t)+v_{j-1}^{x_j+1} F_{k,N+2-j}(x_j+1-\tilde y_k,t) \frac{v_j}{v_{j-1}}. \nonumber
\end{eqnarray}
Using this identity in the previous formula, the first term cancels being proportional to its left column, and the second term yields
\begin{eqnarray}\label{eqDyn3}
G(\ldots,x_j,x_{j-1}=x_j,\ldots;t) = \frac{v_j}{v_{j-1}} G(\ldots,x_j,x_{j-1}=x_j+1,\ldots;t).
\end{eqnarray}

With (\ref{eqDyn3}) we can go back to (\ref{eqDyn1}). First, consider all the terms in (\ref{eqDyn1}) which are proportional to $\dot{a}(t)$. They are given by
\begin{eqnarray}
\hspace{-3em}& & - \sum_{k=1}^N v_k G(\ldots;t) + \sum_{k=1}^N v_k G(\ldots,x_k-1,\ldots;t) \label{eq4.11a}\\
\hspace{-3em}&=& - v_1 G(\ldots;t)-\sum_{k=2}^N v_k(1-\delta_{x_{k-1},x_k+1}) G(\ldots;t) \label{eq4.11b} \\
\hspace{-3em}& & + v_N G(x_N-1,\ldots;t) + \sum_{k=1}^{N-1} v_k(1-\delta_{x_{k+1},x_k}) G(\ldots,x_k-1,\ldots;t) \label{eq4.11c} \\
\hspace{-3em}& & - \sum_{k=2}^N v_k G(\ldots,x_k,x_{k-1}=x_k+1,\ldots;t) \label{eq4.11d} \\
\hspace{-3em}& & + \sum_{k=1}^{N-1} v_k G(\ldots,x_{k+1}=x_k,x_k,\ldots;t). \label{eq4.11e}
\end{eqnarray}
The notation means that the first term of (\ref{eq4.11a}) has been
subdivided into (\ref{eq4.11b}), which contains non-zero terms when
$x_{k-1}\neq x_k+1$, and (\ref{eq4.11d}), whose terms are non-zero
only when $x_{k-1}=x_k+1$. Similarly for the second term of
(\ref{eq4.11a}). By using (\ref{eqDyn3}) and shifting the summation
index by one, we get that (\ref{eq4.11e}) equals
\begin{equation}
\sum_{k=2}^N v_{k-1} G(\ldots,x_k,x_{k-1}=x_k+1,\ldots;t)\frac{v_k}{v_{k-1}},
\end{equation}
which cancels (\ref{eq4.11d}). The expression (\ref{eq4.11b}) is the contribution in the master equation of the particles jumping to the right and leaving the state $(x_N,\ldots,x_1)$ with jump rate $\dot{a}(t) v_k$, while (\ref{eq4.11c}) is the contribution of the particles arriving to the state $(x_N,\ldots,x_1)$. Therefore, the jumps to the right satisfy the exclusion constraint.

Secondly, consider all the terms in (\ref{eqDyn1}) which are proportional to $\dot{b}(t)$. They are
\begin{equation}\label{eqDyn5}
-\sum_{k=1}^N \frac{1}{v_k} G(\ldots;t) + \sum_{k=1}^N \frac{1}{v_k} G(\ldots,x_k+1,\ldots;t).
\end{equation}
Let us denote by $m(k)$ the index of the last particle to the right of particle $k$ such that particle $m(k)$ belongs to the same block of particles as particle $k$ (we say that two particles are in the same block if between them all sites are occupied). Then, (\ref{eqDyn5}) takes the form
\begin{equation}
(\ref{eqDyn5})=-\sum_{k=1}^N \frac{1}{v_k} G(\ldots;t)+\sum_{k=1}^N \frac{1}{v_k}G(\ldots,x_k+1,x_k+1,\ldots,x_k+k-m(k),\ldots;t).
\end{equation}
Using (\ref{eqDyn3}) we get
\begin{eqnarray}
& &\frac{1}{v_k}G(\ldots,x_k+1,x_k+1,\ldots,x_k+k-m(k),\ldots;t)\nonumber \\
&=&\frac{1}{v_k} \frac{v_k}{v_{k-1}} G(\ldots,x_k+1,x_{k}+2,\ldots,x_k+k-m(k),\ldots;t)\\
&=&\frac{1}{v_{k-1}} G(\ldots,x_k+1,x_{k-1}+1,\ldots,x_k+k-m(k),\ldots;t).
\end{eqnarray}
By iterations we finally obtain
\begin{equation}\label{eqDyn6}
(\ref{eqDyn5})=-\sum_{k=1}^N \frac{1}{v_k} G(\ldots;t)+\sum_{k=1}^N \frac{1}{v_{m(k)}}G(\ldots,x_k+1,x_{k-1}+1,\ldots,x_{m(k)}+1,\ldots;t).
\end{equation}
The first term in (\ref{eqDyn6}) is the contribution of particles pushing to the left and leaving the state $(x_N,\ldots,x_1)$, while the second term is the contribution of particles arriving at the state $(x_N,\ldots,x_1)$ because they were pushed, and the particle number $k$ pushes to the left with rate $\dot{b}(t)/v_k$.
\end{proofOF}

We would like to obtain the joint distribution of particle $N_k$ at time $t_k$ for $N_1\geq N_2\geq \ldots \geq N_m\geq 1$ and $0\leq t_1\leq t_2 \leq \ldots \leq t_m$. By Proposition~\ref{Prop1}, this can be written as an appropriate marginal of a product of $m$ determinants (by summing over all variables except the $x_1^{N_k}(t_k)$, $k=1,\ldots,N_m$ under consideration).

\textbf{Notational remark:} Below there is an abuse of notation. For example, $x_l^n(t_i)$ and $x_l^n(t_{i+1})$ are considered different variables even if $t_i=t_{i+1}$. One could call them simply $x_l^n(i)$ and $x_l^n(i+1)$, but then one loses the connection with the times $t_i$'s. In this sense, $t_i$ is considered as a symbol, not as a number.

\begin{thm}\label{ThmDetMeasure}
Let us set $t_0=0$, $a(t_0)=b(t_0)=0$, and $N_{m+1}=0$. The joint
distribution of PushASEP particles is a marginal of a (generally
speaking, signed) measure, obtained by summation of the variables in
the set
\begin{equation}
D=\{x_k^l(t_i),1\leq k \leq l, 1\leq l \leq N_i, 0\leq i \leq m\}\setminus\{x_1^{N_i}(t_i),1\leq i \leq m\};
\end{equation}
the range of summation for any variable in this set in $\Z$. Precisely,
\begin{eqnarray}\label{eqDetMeas}
& & \hspace{-0.5cm}\Pb(x_{N_i}(t_i)=x_1^{N_i}(t_i),1\leq i \leq m | x_k(0)=y_k,1\leq k \leq N_1)\nonumber \\
&& = \cte \times \sum_{D} \det\left[\Psi^{N_1}_{N_1-l}(x_k^{N_1}(t_0))\right]_{1\leq k,l \leq N_1} \nonumber \\
& &\times \prod_{i=1}^{m}\Bigg[\det[{\cal T}_{t_i,t_{i-1}}(x_l^{N_i}(t_i),x_k^{N_i}(t_{i-1}))]_{1\leq k,l \leq N_i}\nonumber \\
& &\phantom{\prod_{i=1}^{m}\Big(}\times \prod_{n=N_{i+1}+1}^{N_i} \det[\phi_n(x_k^{n-1}(t_i),x_l^n(t_i))]_{1\leq k,l\leq n}\Bigg]
\end{eqnarray}
where
\begin{eqnarray}
{\cal T}_{t_j,t_i}(x,y)&=&\frac{1}{2\pi\I}\oint_{\Gamma_0}\dx z z^{x-y-1} e^{(a(t_j)-a(t_i))/z} e^{(b(t_j)-b(t_i))z},\label{eqW} \\
\Psi^{N_1}_{N_1-l}(x)&=&\frac{1}{2\pi\I}\oint_{\Gamma_0}\dx z z^{x-y_l-1} (1-v_{l+1} z)\cdots (1-v_{N_1}z), \label{eqPsi}\\
\phi_n(x,y)&=&v_n^{y-x}\Id_{[y\geq x]}\quad \textrm{ and }\quad \phi_n(x_n^{n-1},y)=v_n^y.\label{eqphi}
\end{eqnarray}
The normalizing constant in (\ref{eqDetMeas}) is chosen so that the
sum over all variables $\{x_k^l(t_i),\,1\leq k \leq l,\, 1\leq l
\leq N_i,\, 0\leq i \leq m\}$ equals 1.
\end{thm}
\textbf{Remark:} the variables $x_n^{n-1}$ participating in the last factor of (\ref{eqDetMeas}) are fictitious, cf.\ (\ref{eqphi}), and are used for convenience of notation only.

\begin{figure}[t!]
\begin{center}
\psfrag{0}[r]{$0$}
\psfrag{n2}[r]{$N_2$}
\psfrag{n1}[r]{$N_1$}
\psfrag{t0}[c]{$t_0$}
\psfrag{t1}[c]{$t_1$}
\psfrag{t2}[c]{$t_2$}
\psfrag{112}[r]{$x_1^1(t_2)$}
\psfrag{1n22}[r]{$x_1^{N_2}(t_2)$}
\psfrag{n2n22}{$x_{N_2}^{N_2}(t_2)$}
\psfrag{1n21}[r]{$x_1^{N_2}(t_1)$}
\psfrag{n2n21}{$x_{N_2}^{N_2}(t_1)$}
\psfrag{1n11}[r]{$x_1^{N_1}(t_1)$}
\psfrag{n1n11}{$x_{N_1}^{N_1}(t_1)$}
\psfrag{1n10}[r]{$x_{1}^{N_1}(t_0)$}
\psfrag{n1n10}{$x_{N_1}^{N_1}(t_0)$}
\includegraphics[height=6cm]{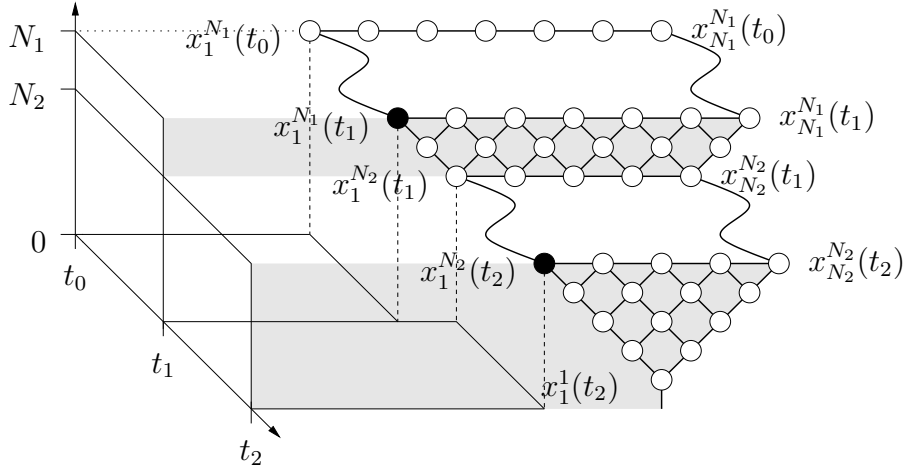}
\caption{A graphical representation of variables entering in the determinantal structure, illustrated for $m=2$. The wavy lines represents the time evolution between $t_0$ and $t_1$ and from $t_1$ to $t_2$. The rest is the interlacing structure on the variables induced by the $\det[\phi_n(\cdots)]$. The black dots are the only variables which are not in the summation set $D=D(0)\cup D^*(t_1)\cup \cdots \cup D^*(t_m)$ (see Figure~\ref{FigDomains} too). The variables of the border of the interlacing structures are explicitly indicated.}
\label{FigDetStructureB}
\end{center}
\end{figure}

We illustrate the determinantal structure in Figure~\ref{FigDetStructureB}.

\begin{proofOF}{Theorem~\ref{ThmDetMeasure}}
Since the evolution is Markovian, we have
\begin{eqnarray}\label{eq2.32}
& & \Pb(x_{N_i}(t_i)=x_1^{N_i}(t_i),1\leq i \leq m | x_k(0)=y_k,1\leq k \leq N_1)\nonumber \\
&=& \sum \Id(x_1^k(0)=y_k,1\leq k \leq N_1) \\
&\times & \prod_{i=1}^m\Pb(x_k(t_i)=x_1^k(t_i),1\leq k \leq N_i | x_k(t_{i-1})=x_1^k(t_{i-1}),1\leq k \leq N_i) \nonumber
\end{eqnarray}
where the sum is over $x_1^k(0)$, $1\leq k \leq N_1$, and $x_1^k(t_i)$, $1\leq k \leq N_i-1$, $i=1,\ldots,m$. Note that so far the lower index of all variables $x_l^k$ is identically equal to $1$.

The continuation of the proof requires a series of Lemmas collected
at the end of this section, see Section~\ref{SectLemmas}. We apply
Proposition~\ref{Prop1} to the $m+1$ factors in (\ref{eq2.32})
(including the indicator function, which corresponds to the value
$t=0$ in Proposition~\ref{Prop1}). Namely,
\begin{eqnarray}\label{eq2.33}
& &\hspace{-1.5em} \Pb(x_k(t_i)=x_1^k(t_i),1\leq k \leq N_i | x_k(t_{i-1})=x_1^k(t_{i-1}),1\leq k \leq N_i)  \\
&&\hspace{-1.5em}= \cte \times \Big(\prod_{n=1}^{N_i} v_n^{x_1^n(t_i)-x_1^n(t_{i-1})}\Big) \det\left[F_{k,l}(x_1^{N_i+1-l}(t_i)-x_1^{N_i+1-k}(t_{i-1}),a_i,b_i)\right]_{1\leq k,l\leq N_i},\nonumber
\end{eqnarray}
where we introduced the notation $a_i:=a(t_i)-a(t_{i-1})$, and $b_i:=b(t_i)-b(t_{i-1})$.

First we collect all the factors coming from the $\prod_{n=1}^{N_i} v_n^{x_1^n(t_i)-x_1^n(t_{i-1})}$. We have the factor
\begin{eqnarray}\label{eq2.34}
& &\Big(\prod_{n=1}^{N_1} v_n^{x_1^n(0)-y_n}\Big)\prod_{k=1}^m \prod_{n=1}^{N_n} v_n^{x_1^n(t_k)-x_1^n(t_{k-1})} \nonumber \\
&=& \Big(\prod_{n=1}^{N_1} v_n^{-y_n}\Big) \Big(\prod_{i=1}^{m-1} \prod_{n=N_{i+1}+1}^{N_i} v_n^{x_1^n(t_i)}\Big) \prod_{n=1}^{N_m}v_n^{x_1^n(t_m)}.
\end{eqnarray}
Then we apply Lemma~\ref{LemmaExpand} to all the factors $\det[F_{k,l}(\cdots)]$. For the initial condition we have
\begin{equation}\label{eq2.35}
\sum_{\widetilde D(0)} \det\left[F_{N_1+1-l,1}(x_k^{N_1}(0)-y_l,0,0)\right]_{1\leq k,l\leq N_1}
\prod_{n=2}^{N_1} \det\left[\varphi_n(x_k^{n-1}(0),x_l^n(0))\right]_{1\leq k,l \leq n}.
\end{equation}
For the other terms, $i=1,\ldots,m$, we get
\begin{eqnarray}\label{eq2.36}
& &\sum_{\widetilde D(t_i)} \det\left[F_{N_i+1-l,1}(x_k^{N_i}(t_i)-x_1^l(t_{i-1}),a_i,b_i)\right]_{1\leq k,l\leq N_i}
\nonumber \\
& &\hspace{2em}\times\prod_{n=2}^{N_i} \det\left[\varphi_n(x_k^{n-1}(t_i),x_l^n(t_i))\right]_{1\leq k,l \leq n}.
\end{eqnarray}

Thus, the probability we want to compute in (\ref{eq2.32}) is obtained by a marginal of a measure on $m+1$ interlacing triangles, when we sum over all the variables in $D(0), D^*(t_1), \ldots, D^*(t_m)$, see Figure~\ref{FigDomains} for the definitions of these sets. At this point we apply Lemma~\ref{LemmaContract} as follows. For $i=1,\ldots,m-1$ we do the sum over the variables in $\widehat D(t_i)$. Notice that the remaining variables in (\ref{eq2.34}) do not belong to the $\widehat D(t_i)$, thus we factorize them out. So, r.h.s.\ of (\ref{eq2.32}) is, up to a constant, equal to
\begin{eqnarray}\label{eq2.37}
& &\sum (\ref{eq2.34})\times \det\left[F_{N_1+1-l,1}(x_k^{N_1}(0)-y_l,0,0)\right]_{1\leq k,l\leq N_1}\nonumber \\
&\times & \Bigg[\prod_{i=0}^{m-1}\Big(\prod_{n=2}^{N_i} \det\left[\varphi_n(x_k^{n-1}(t_i),x_l^n(t_i))\right]_{1\leq k,l \leq n}\Big)\nonumber \\
&\times &\det\left[F_{N_{i+1}+1-l,1}(x_k^{N_{i+1}}(t_{i+1})-x_1^l(t_{i}),a_{i+1},b_{i+1})\right]_{1\leq k,l\leq N_{i+1}}\Bigg]\nonumber \\
&\times & \prod_{n=2}^{N_m} \det\left[\varphi_n(x_k^{n-1}(t_m),x_l^n(t_m))\right]_{1\leq k,l \leq n}
\end{eqnarray}
with the sum is over the variables described just above. By summing over the $\widehat D(t_i)$, the determinant with $F_{N_{i+1}+1-l,1}$ becomes a determinant with $F_{1,1}$ and the product of the $\det[\varphi_n(\cdots)]$ is restricted to $n=N_{i+1}+1,\ldots,N_i$. Thus,
\begin{eqnarray}
(\ref{eq2.32})&=&\cte \times \sum (\ref{eq2.34})\times \det\left[F_{N_1+1-l,1}(x_k^{N_1}(0)-y_l,0,0)\right]_{1\leq k,l\leq N_1}\nonumber \\
&\times & \prod_{i=1}^{m} \Big(\det\left[F_{1,1}(x_k^{N_{i}}(t_{i})-x_l^{N_{i}}(t_{i-1}),a_{i},b_{i})\right]_{1\leq k,l\leq N_i}\nonumber \\
& &\times\prod_{n=N_{i+1}+1}^{N_i} \det\left[\varphi_n(x_k^{n-1}(t_i),x_l^n(t_i))\right]_{1\leq k,l \leq n}\Big)
\end{eqnarray}
where we set $N_{m+1}=0$ (the contribution from $n=1$ is $1$). Finally, by using Lemma~\ref{LemPhis} we can include the terms in (\ref{eq2.34}) into the $\varphi_n$'s by modifying the last row, i.e., by setting it equal to $v_n^y$. Thus,
\begin{eqnarray}
(\ref{eq2.32})&=&\cte \times \det\left[F_{N_1+1-l,1}(x_k^{N_1}(0)-y_l,0,0)\right]_{1\leq k,l\leq N_1}\nonumber \\
&\times & \prod_{i=1}^{m} \Big(\det\left[F_{1,1}(x_k^{N_{i}}(t_{i})-x_l^{N_{i}}(t_{i-1}),a_{i},b_{i})\right]_{1\leq k,l\leq N_i}\nonumber \\
& &\times\prod_{n=N_{i+1}+1}^{N_i} \det\left[\phi_n(x_k^{n-1}(t_i),x_l^n(t_i))\right]_{1\leq k,l \leq n}\Big).
\end{eqnarray}
The identification to the expressions in Theorem~\ref{ThmDetMeasure}
uses the representations (\ref{eqProp1b}) and (\ref{eqPSI}).
\end{proofOF}

The first line represent the initial condition at $t_0=0$, the term with $\Psi^{N_1}_{N_1-l}$ in Theorem~\ref{ThmDetMeasure}. These $N_1$ variables evolves until time $t_1$ and this is represented by the first line (term ${\cal T}_{t_1,t_0}$). After that, there is a reduction of the number of variables from $N_1$ to $N_2$ by the interlacing structure, which is followed by the time evolution from $t_1$ to $t_2$. This is repeated $m-1$ times. Finally it ends with an interlacing structure. If $N_1=N_2$, then the first interlacing structure is trivial (not present), while if for example $t_2=t_1$, then the time evolution is just the identity.

In what follows, the picture to keep in mind consists of reading
Figure~\ref{FigDetStructureB} from bottom to top, i.e., in the
reversed order with respect to the original decomposition. Then
$n_i$ increases and $t_i$ decreases. This corresponds to having a
sort of vicious walkers with increasing number of walkers when the
transition is made by the $\phi$'s, and with constant number of
walkers if the transition is the temporal one made by ${\cal T}$.

The measure in (\ref{eqDetMeas}) is written with the outer product
over time moments but it can be rewritten by taking the outer
product over the index $n$ in the variables $x_k^n$'s. Let us
introduce the following notations. For any level $n$ there is a
number $c(n)\in\{0,\ldots,m+1\}$ of products of terms ${\cal T}$
which are the time evolution of $n$ particles between consecutive
times in the set $\{t_1,\ldots,t_m\}$ (in other words $c(n)$ is
$\#\{i|N_i=n\}$). Let us denote them by $t^n_0<\ldots<t^n_{c(n)}$.
Notice that $t^n_{0}=t^{n+1}_{c(n+1)}$, $t^{N_1}_0=t_0$,
$t^{N_1}_1=t_1$, and $t_0^0=t_{c(0)}^0=t_m$. Then, the measure in
(\ref{eqDetMeas}) takes the form
\begin{eqnarray}\label{eqDetMeasB}
& &\hspace{-1em}\cte\times \prod_{n=1}^{N_1} \Bigg[\det[\phi_n(x_k^{n-1}(t_0^{n-1}),x_l^n(t^n_{c(n)}))]_{1\leq k,l\leq  n} \\
& & \times\prod_{a=1}^{c(n)} \det[{\cal T}_{t_a^n,t_{a-1}^n}(x_k^n(t^n_a),x^n_l(t^n_{a-1}))]_{1\leq k,l \leq n}
\Bigg]\det[\Psi^{N_1}_{N_1-l}(x^{N_1}_k(t_0^{N_1}))]_{1\leq k,l \leq N_1}. \nonumber
\end{eqnarray}

In Theorem~\ref{ThmKernel} we show that a measure on the $x^n_k(t^n_a)$ of the form (\ref{eqDetMeasB}) is determinantal and we give the expression for the kernel. Then we particularize it in case of the PushASEP with particle dependent jump rates. For this purpose, we introduce a couple of notations. For any two time moments $t_{a_1}^{n_1},t_{a_2}^{n_2}$, we define the convolution over all the transitions between them by $\phi^{(t_{a_1}^{n_1},t_{a_2}^{n_2})}$ (backwards in time, since forward in the $n$'s).
For $(n_1,t_1)\prec (n_2,t_2)$ (see the definition in (\ref{DefPrec})), we set
\begin{equation}\label{eq4.32}
\phi^{(t_{a_1}^{n_1},t_{a_2}^{n_2})}={\cal T}_{t^{n_1}_{a_1},t^{n_1}_{0}} * \, \phi_{n_1+1}*{\cal T}^{n_1+1} *\cdots*\phi_{n_2}*{\cal T}_{t^{n_2}_{c(n_2)},t^{n_2}_{a_2}}
\end{equation}
where
\begin{equation}
{\cal T}^n = {\cal T}_{t_{c(n)}^n,t_0^n}.
\end{equation}
For $(n_1,t_1)\not\prec (n_2,t_2)$ we set
$\phi^{(t_{a_1}^{n_1},t_{a_2}^{n_2})}=0$. Above we used
\begin{equation}\label{eq2.12}
{\cal T}_{t_3,t_2}*{\cal T}_{t_2,t_1} = {\cal T}_{t_3,t_1},
\end{equation}
which is an immediate corollary of (\ref{eqW}). In a more general case considered in Theorem~\ref{ThmKernel} below, if (\ref{eq2.12}) does not holds, then ${\cal T}^n$ is just the convolution of the transitions between $t_{c(n)}^n$ and $t_0^n$ by definition.
Moreover, define the matrix $M$ with entries $M_{k,l}$, $1\leq k,l\leq N_1$,
\begin{equation}
M_{k,l}=\big(\phi_{k}*{\cal T}^k*\cdots *\phi_{N_1}*{\cal T}^{N_1}*\Psi^{N_1}_{N_1-l}\big)(x_k^{k-1})
\end{equation}
and the vector
\begin{equation}
\Psi^{n,t^n_a}_{n-l}=\phi^{(t^n_a,t^{N_1}_0)}*\Psi^{N_1}_{N_1-l}.
\end{equation}
We remind that the variables $x_k^{k-1}$ in $M_{k,l}$ are fictitious, compare with (\ref{eqphi}).

\begin{thm}\label{ThmKernel}
Assume that the matrix $M$ is invertible. Then the normalizing
constant in (\ref{eqDetMeasB}) is equal to $(\det M)^{-1}$, the
normalized measure\footnote{With \emph{normalized measure} we mean that all weights add up to one. If all weights are non-negative, it is a \emph{probability measure} (this is the case for example for PushASEP with step initial conditions).} of the form (\ref{eqDetMeasB}) viewed as
$(N_1+\ldots+N_m)$-point process is determinantal, and the
correlation kernel can be computed as follows
\begin{eqnarray}
K(t^{n_1}_{a_1},x_1; t^{n_2}_{a_2},x_2)&=& -\phi^{(t^{n_1}_{a_1},t^{n_2}_{a_2})}(x_1,x_2) \\
&+& \sum_{k=1}^{N_1} \sum_{l=1}^{n_2} \Psi^{n_1,t^{n_1}_{a_1}}_{n_1-k}(x_1) [M^{-1}]_{k,l} (\phi_l * \phi^{(t^l_{c(l)},t^{n_2}_{a_2})})(x^{l-1}_l,x_2). \nonumber
\end{eqnarray}
In the case when the matrix $M$ is upper triangular, there is a simpler way to write the kernel.
Set
\begin{equation}\label{eq4.37}
\Phi^{n,t_a^n}_{n-k}(x)=\sum_{l=1}^n [M^{-1}]_{k,l} \big(\phi_l * \phi^{(t^l_{c(l)},t^n_a)}\big)(x^{l-1}_l,x)
\end{equation}
for all $n=1,\ldots,N_1$ and $k=1,\ldots,n$. Then, $\big\{\Phi^{n,t^n_a}_{n-k}\big\}_{k=1,\ldots,n}$ is the unique basis of the linear span of
\begin{equation}\label{eq2.16}
\Big\{(\phi_1* \phi^{(t^1_{c(1)},t^{n}_a)})(x_1^0,x), \ldots, (\phi_{n}*\phi^{(t^{n}_{c(n)},t^{n}_a)})(x_{n}^{n-1},x) \Big\}
\end{equation}
that is biorthogonal to $\{\Psi^{n,t^n_a}_{n-k}\}$:
\begin{equation}
\sum_{x\in \Z}\Phi^{n,t^n_a}_i(x) \Psi^{n,t^n_a}_j(x) = \delta_{i,j},\quad i,j=0,\ldots,n-1.
\end{equation}
The correlation kernel can then be written as
\begin{equation}
K(t^{n_1}_{a_1},x_1; t^{n_2}_{a_2},x_2)= -\phi^{(t^{n_1}_{a_1},t^{n_2}_{a_2})}(x_1,x_2) + \sum_{k=1}^{n_2} \Psi^{n_1,t^{n_1}_{a_1}}_{n_1-k}(x_1) \Phi^{n_2,t^{n_2}_{a_2}}_{n_2-k}(x_2).
\end{equation}
Moreover, one has the identity
\begin{equation}
\phi^{(t^{n_1}_{a_1},t^{n_2}_{a_2})} *\Phi^{n_2,t^{n_2}_{a_2}}_{n_2-l}=\Phi^{n_1,t^{n_1}_{a_1}}_{n_1-l}
\end{equation}
for $n_1\geq n_2$ and $a_1\leq a_2$ for $n_1=n_2$.
\end{thm}
\begin{proofOF}{Theorem~\ref{ThmKernel}}
The proof is similar to the one of Lemma 3.4 in~\cite{BFPS06}, which
is in its turn based on the formalism of~\cite{RB04}. The only place
where the argument changes substantially is the definition of the
matrix $L$, see~\cite{BFPS06}, formula (3.32). We need to construct
the matrix $L$ in such a way that its suitable minors reproduce, up
to a common constant, the weights (\ref{eqDetMeasB}) of the measure.
Then our measure turns into a \textit{conditional $L$-ensemble} in
the terminology of~\cite{RB04}.

The variables of interest live in the space $\mathfrak{Y}=\mathfrak{X}^{(1)}\cup
\cdots \cup \mathfrak{X}^{(N_1)}$, with
$\mathfrak{X}^{(n)}=\mathfrak{X}^{(n)}_0\cup\cdots \cup
\mathfrak{X}^{(n)}_{c(n)}$, where $\mathfrak{X}^{(n)}_a=\Z$ is the
space where the $n$ variables at time $t^n_a$ live. Let us also
denote $I=\{1,\ldots,N_1\}$. Then, the matrix $L$ written with the
order given by the entries in the set of all variables
\mbox{$\mathfrak{X}=I \cup \mathfrak{Y}$} becomes
\begin{equation}
L=\left(
  \begin{array}{c@{\hspace{0.1em}}c@{\hspace{0.1em}}c@{\hspace{0.1em}}c@{\hspace{0.1em}}c@{\hspace{0.1em}}c@{\hspace{0.1em}}c@{\hspace{0.1em}}c@{\hspace{0.1em}}c@{\hspace{0.1em}}c}
    0 & E_0 & 0 & E_1 & 0 & E_2 & 0 & \cdots & E_{N_1-1} & 0 \\
    0 & 0 & -T_1 & 0 & 0 & 0 & 0 & \cdots & 0 & 0 \\
    0 & 0 & 0 & -W_{[1,2)} & 0 & 0 & 0 &\cdots & 0 & 0 \\
    0 & 0 & 0 & 0 & -T_2 & 0 & 0 &\cdots& 0 & 0 \\
    0 & 0 & 0 & 0 & 0 & -W_{[2,3)} & 0 &\cdots& 0 & 0 \\
    0 & 0 & 0 & 0 & 0 & 0 & -T_3 &\cdots& 0 & 0\\
    \vdots &  \vdots &  \vdots &  \vdots &  \vdots &  \vdots &  \vdots &\ddots&  \vdots &  \vdots \\
    0 & 0 & 0 & 0 & 0 & 0 & 0 &\cdots& -W_{[N_1-1,N_1)} & 0 \\
    0 & 0 & 0 & 0 & 0 & 0 & 0 &\cdots& 0 & -T_{N_1} \\
    \Psi^{(N_1)} & 0 & 0 & 0 & 0 & 0 & 0 &\cdots& 0 & 0 \\
  \end{array}
\right)
\end{equation}
with the matrix blocks in $L$ have the following entries:
\begin{eqnarray}
[\Psi^{(N_1)}]_{x,j}&=&\Psi^{N_1}_{N_1-j}(x), \quad x\in \mathfrak{X}^{(N_1)}_0,j \in I,\\
\phantom{}[E_n]_{i,y}&=&\begin{cases}
\phi_{n+1}(x^n_{n+1},y),& i=n+1,y\in \mathfrak{X}^{(n+1)}_{c(n+1)},\\
0, &i\in I\setminus\{n+1\},y\in \mathfrak{X}^{(n+1)}_{c(n+1)},
\end{cases} \\
\phantom{}[W_{[n,n+1)}]_{x,y}&=&\phi_{n+1}(x,y),\quad x\in\mathfrak{X}^{(n)}_{0}, y\in \mathfrak{X}^{(n+1)}_{c(n+1)},
\end{eqnarray}
and $T_n$ is the matrix made of blocks
\begin{equation}
T_n=\left(
      \begin{array}{ccc}
        T_{n,1} & 0 & 0 \\
        0 & \ddots & 0 \\
        0 & 0 & T_{n,c(n)}
      \end{array}
    \right),
\end{equation}
where
\begin{equation}
[T_{n,a}]_{x,y}={\cal T}_{t^n_a,t^n_{a-1}}(x,y), \quad x\in \mathfrak{X}^{(n)}_a,y\in\mathfrak{X}^{(n)}_{a-1}.
\end{equation}
The rest of the proof is along the same lines as that of Lemma 3.4 in~\cite{BFPS06}.

Although the argument gives a proof in the case when all variables $x_a^n(t_b^n)$ vary over finite sets, a simple limiting argument immediately extends the statement to any discrete sets, provided the series that defines $M_{k,l}$ are absolutely convergent, which is certainly true in our case.
\end{proofOF}

A special case of Theorem~\ref{ThmKernel} is Proposition~\ref{PropPushASEP} stated in Section~\ref{SectKernels}, which we prove below.
\begin{proofOF}{Proposition~\ref{PropPushASEP}}
We first prove the statement in the case the jump rates are ordered,
$v_1>v_2>\dots$, and then use analytic continuation in $v_j$'s.

For $v_1>v_2>\dots$, the claim is a specialization of
Theorem~\ref{ThmKernel}. The kernel depends only on the actual times
and particle numbers, therefore we might drop the label $a_i$ of
$t^{n_i}_{a_i}$. Equivalently, we can use the notation $(n_i,t_i)$
instead of $t^{n_i}_{a_i}$, to go back to the natural notations of
the model. For PushASEP we have
$\Psi^{N_1}_{N_1-l}(x)=F_{N_1+1-l,1}(x-y_l,0,0)$ and
\begin{equation}
{\cal T}_{t_j,t_i}(x,y)=F_{1,1}(x-y,a(t_j)-a(t_i),b(t_j)-b(t_i)).
\end{equation}
First of all, we sum over the $\{x_k^{N_1}(0),1\leq k \leq N_1\}$ variables, since we are not interested in the initial conditions (being fixed). When applied to the $F_{k,l}(x,a(t_i),b(t_i))$, the time evolution ${\cal T}_{t_j,t_i}$ changes it into $F_{k,l}(x,a(t_j),b(t_j))$,
\begin{equation}\label{eqP1}
\sum_{y\in\Z} {\cal T}_{t_j,t_i}(x,y) F_{k,l}(y,a(t_i),b(t_i))=F_{k,l}(x,a(t_j),b(t_j)).
\end{equation}
This implies that Theorem~\ref{ThmKernel} still holds but with $t_0^{N_1}=t_1$ and
\begin{equation}
\Psi^{N_1}_{N_1-l}(x)=F_{N_1+1-l,1}(x-y_l,a(t_1),b(t_1)).
\end{equation}
We have, see (\ref{eqRecRel1}), that
\begin{equation}\label{eqP2}
(\phi_k * F_{l,N_1+1-k})(x,a,b)=F_{l,N_1+2-k}(x,a,b).
\end{equation}
Using (\ref{eqP1}) and (\ref{eqP2}) repeatedly one then gets
\begin{equation}
\Psi^{n,t^n_k}_{n-l}(x) = F_{N_1+1-l,N_1+1-n}(x-y_l,a(t^n_k),b(t^n_k))
\end{equation}
which can be rewritten as (\ref{eqPSI}).

Next we show that the matrix $M$ is upper triangular. Once again, (\ref{eqP1}) and (\ref{eqP2}) are applied several times, leading to
\begin{equation}
M_{k,l}=\sum_{y\in\Z}v_k^y F_{N_1+1-l,N_1+1-k}(y-y_l,a(t^k_{c(k)}),b(t^k_{c(k)})).
\end{equation}
Set $a_k=a(t^k_{c(k)})$ and $b_k=b(t^k_{c(k)})$. Then,
\begin{equation}
M_{k,l}=\sum_{y\in\Z} v_k^y \frac{1}{2\pi\I}\oint_{\Gamma_0} \dx z z^{y-y_l-1} e^{a_k/z+b_k z} \frac{(1-v_{l+1}z)\cdots (1-v_{N_1}z)}{(1-v_{k+1}z)\cdots (1-v_{N_1}z)}.
\end{equation}
(Note that we need the assumption $v_k>\max\{v_l\}_{l>k}$ in order
for this sum to converge.) We divide the sum over $y$ in two
regions, $\{y\geq 0\}$ and $\{y<0\}$. The sum over $y\geq 0$ can be
taken into the integral provided that $|v_k z|<1$ and then we use
$\sum_{y\geq 0} (v_k z)^y=\frac{1}{1-v_k z}$. Similarly, the sum
over $y<0$ is taken into the integrals provided that $|v_k z|>1$ and
one uses $\sum_{y\geq 0} (v_k z)^y=-\frac{1}{1-v_k z}$. For $k>l$
the new term in the denominator, $1-v_k z$, is canceled so that
this is not a pole and we can deform the contours to be the same.
Thus for $k>l$ the net result is zero. This is not the case for
$k\leq l$, since in that case the new pole at $1/v_k$ does not have
to vanish. The diagonal term is easy to compute, since the pole at
$1/v_k$ is simple. Computing its residue we obtain $M_{k,k}=
v_k^{y_l+1} e^{v_k a_k+b_k/v_k}$ and
\begin{equation}\label{detM}
\det M=\prod_{k=1}^{N_1} v_k^{y_l+1} e^{v_k a_k+b_k/v_k}\ne 0.
\end{equation}

Next, we need to determine the space $V_{n}$ where the
orthogonalization has to be made. We have
\begin{equation}
(\phi_{k}*\phi^{(t^{k}_{c(k)},t_1)})(x^{k-1}_k,x)=
\sum_{y\in\Z}v_{k}^y \frac{1}{2\pi\I}
\oint_{\Gamma_0}\dx z z^{y-x-1} \frac{e^{a_k/z+b_k z}}{(1-v_{k+1}z)\cdots(1-v_{N_1}z)}.
\end{equation}
Once again we divide the sum over $\{y\geq 0\}$, $\{y<0\}$ and then
deform the paths so that the only remaining contribution is the
residue at $z=1/v_k$, which is equal to $const\cdot v_k^x$. Thus, $V_n={\rm span}(v_1^x,\ldots,v_n^x)$, $n=1,\ldots,N_1$.

Finally, we need an expression for the transition between two times,
which is given by (\ref{eq4.32}). Every time that we convolute a
$\phi_k$ with $\cal T$, we get an extra factor $1/(1-v_k z)$ in the
integral. Therefore, if $t^{n_2}_{a_2}\leq t^{n_1}_{a_1}$ and
$n_2\geq n_1$, then
\begin{equation}
\phi^{(t^{n_1}_{a_1},t^{n_2}_{a_2})}(x,y)=
\frac{1}{2\pi\I}\oint_{\Gamma_0}\dx z
z^{x-y-1}\frac{e^{(a(t^{n_1}_{a_1})-a(t^{n_2}_{a_2}))/z}e^{(b(t^{n_1}_{a_1})-b(t^{n_2}_{a_2}))z}}{(1-v_{n_1+1}z)\cdots
(1-v_{n_2}z)},
\end{equation}
while $\phi^{(t^{n_1}_{a_1},t^{n_2}_{a_2})}(x,y)=0$ otherwise.

Now let us do the analytic continuation.

As all the functions $\Psi^{n,t}_k$, see (\ref{eqPSI}), can be
estimated as
\begin{equation}
|\Psi^{n,t}_k(x)|\leq \cte \cdot q^{|x|}, \qquad x\in \Z,
\end{equation}
for any $q>0$ and $v_1,v_2,\dots$ varying in a compact set, the
weights (\ref{eqDetMeasB}) can be majorated  by a convergent series
for $v_1,v_2,\dots$ varying in a compact set. Further, the
normalizing constant $(\det M)^{-1}$ is analytic as long as $v_j$'s
are nonzero, see (\ref{detM}). Thus, the correlation functions of
our measure are analytic in $v_j$'s.

Set, for $k=0,\dots,n-1$,
\begin{equation}\label{eqfk}
f_k(x)=\frac{1}{2\pi\I}\oint \frac{\dx z\, z^{-x-1}}{(1-v_{n-k}z)(1-v_{n-k+1}z)\cdots(1-v_{n}z)},
\end{equation}
where the integration contour includes the poles
$v_{n-k}^{-1},\dots,v_n^{-1}$. Note that $f_k(x)$ is a linear
combination of $v_{n-k}^x,\dots,v_n^x$. Denote by
$G={[G_{k,l}]}_{k,l=0,\ldots,n-1}$ the Gram matrix
\begin{equation}
G_{k,l}=\sum_{x\in\Z} f_k(x)\Psi_l^{n,t}(x).
\end{equation}
Then for $v_1>v_2>\dots$ we have
\begin{equation}\label{eqPhian}
\Phi_k^{n,t}(x)=\sum_{l=0}^{n-1}{[G^{-1}]}_{k,l}f_l(x).
\end{equation}
Since the matrix $M$ is triangular, $G$ is also triangular. Its
diagonal elements are easy to compute:
$G_{k,k}=e^{a(t)v_k+b(t)/v_k}v_k^{y_k+1}$. Hence, (\ref{eqPhian})
gives a formula for $\Phi$'s that is analytic in $v_j$'s as long as
they stay away from zero. This implies that the corresponding
expression for the correlation kernel (\ref{3.4}) is also analytic
in $v_j$'s, and thus both sides of the determinantal formula for the
correlation functions can be analytically continued.
Finally, it is not difficult to see that the functions (\ref{eqfk}) span the
space $V_n$ given by (\ref{eqVn}), which implies the statement.
\end{proofOF}

\subsection{Some lemmas}\label{SectLemmas}
In this subsection we state and prove the Lemmas used in the proof of Theorem~\ref{ThmKernel}.

\begin{lem}\label{LemRecRel} Let us define the function
\begin{equation}
\varphi_n(x,y)=\left\{\begin{array}{ll}
v_n^{y-x},& y\geq x,\\
0,& y<x.
\end{array}\right.
\end{equation}
Then the following recurrence relations holds
\begin{equation}\label{eqRecRel1}
F_{k,l+1}(x,a,b)=(\varphi_{N+1-l}* F_{k,l})(x,a,b)
\end{equation}
and
\begin{equation}\label{eqRecRel2}
F_{k-1,l}(x,a,b)=(\varphi_{N+2-k}*F_{k,l})(x,a,b).
\end{equation}
From (\ref{eqRecRel2}) and $\varphi_n(x,y)=\varphi_n(0,y-x)=\varphi_n(-y,-x)$ it follows
\begin{equation}\label{eqRecRel3}
F_{k-1,l}(-x,a,b)=\sum_{y\in\Z} F_{k,l}(-y,a,b) \varphi_{N+2-k}(y,x).
\end{equation}
\end{lem}
\begin{proofOF}{Lemma~\ref{LemRecRel}}
We have
\begin{equation}
F_{k,l}(x,a,b)=\frac{1}{2\pi \I}\oint_{\Gamma_0} \dx z z^{x-1} e^{b z} e^{a/z}
\frac{(1-v_N z)\cdots (1-v_{N+2-k} z)}{(1-v_N z)\cdots (1-v_{N+2-l} z)}.
\end{equation}
Then applying $\sum_{y\geq x} v_{N+1-l}^{y-x} z^y = z^x/(1-v_{N+1-l}z)$ (for $|z|\ll 1$), we get that in the denominator we have an extra factor, which corresponds to increasing $l$ by one. Similarly, applying $\varphi_{N+2-k}$, the extra factor in the denominator cancels the last one in the numerator, thus this is equivalent to decreasing $k$ by one.
\end{proofOF}

\begin{figure}
\begin{center}
\psfrag{11}{$x_1^1(t_i)$}
\psfrag{22}{$x_2^2(t_i)$}
\psfrag{n-1n-1}{$x_{N_{i+1}-1}^{N_{i+1}-1}(t_i)$}
\psfrag{nn}{$x_{N_{i+1}}^{N_{i+1}}(t_i)$}
\psfrag{mm}{$x_{N_{i}}^{N_{i}}(t_i)$}
\psfrag{1n-1}[r]{$x_{1}^{N_{i+1}-1}(t_i)$}
\psfrag{1n}[r]{$x_{1}^{N_{i+1}}(t_i)$}
\psfrag{1m-1}[r]{$x_{1}^{N_{i}-1}(t_i)$}
\psfrag{1m}{$x_{1}^{N_{i}}(t_i)$}
\psfrag{2m}{$x_{2}^{N_{i}}(t_i)$}
\psfrag{D}{$D(t_i)$}
\psfrag{D*}{$D^*(t_i)$}
\psfrag{Dtilde}{$\widetilde D(t_i)$}
\psfrag{Dhat}{$\widehat D(t_i)$}
\psfrag{D*hat}{$\widehat D^*(t_i)$}
\includegraphics[height=7cm]{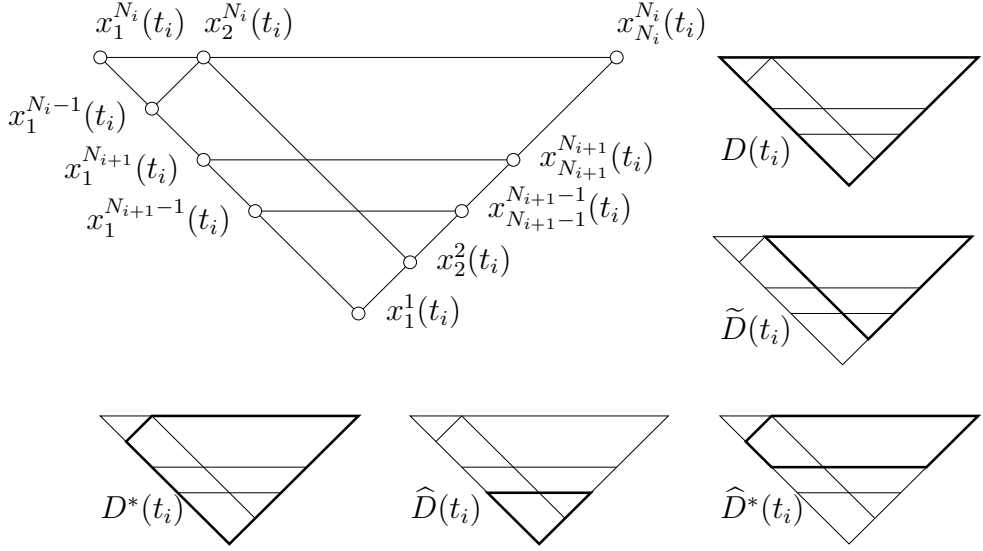}
\caption{A graphical representation of the summation domains that occurs in the next lemmas and theorem. The bold lines passes through the border of the domains.}
\label{FigDomains}
\end{center}
\end{figure}

We define the following domains, which will occurs several times in the following. A graphical representation is in Figure~\ref{FigDomains}.
Let us denote the set of interlacing variables at time $t_i$ by
\begin{equation}
D(t_i)=\{x_k^n(t_i), 1\leq n \leq N_i, 1\leq k \leq n | x_k^{n+1}(t_i)< x_k^n(t_i) \leq x_{k+1}^{n+1}(t_i)\}.
\end{equation}
Then let
\begin{equation}
\widetilde D(t_i)=\{x_k^n(t_i) \in D(t_i) | k\geq 2\},\quad \widehat D(t_i)=\{x_k^n(t_i) \in D(t_i) | n\leq N_{i+1}-1\},
\end{equation}
and
\begin{equation}
D^*(t_i)=D(t_i)\setminus\{x_1^{N_i}(t_i)\},\quad \widehat D^*(t_i)=D^*(t_i)\setminus \widehat D(t_i).
\end{equation}

\begin{lem}\label{LemmaExpand}
We have the identity
\begin{eqnarray}\label{eq2.29}
& &\det\left[F_{k,l}(x_1^{N_i+1-l}(t_i)-x_1^{N_i+1-k}(t_{i-1}),a,b)\right]_{1\leq k,l\leq N_i} \nonumber \\
&=&\cte \sum_{\widetilde D(t_i)} \Big(\prod_{n=2}^{N_i}
\det\left[\varphi_n(x_k^{n-1}(t_i),x_l^n(t_i))\right]_{1\leq k,l \leq n}\Big) \nonumber \\
&\times & \det\left[F_{N_i+1-l,1}(x_k^{N_i}(t_i)-x_1^{l}(t_{i-1}),a,b)\right]_{1\leq k,l\leq N_i}
\end{eqnarray}
where we set $\varphi_n(x_n^{n-1},x)=1$.
\end{lem}
\begin{proofOF}{Lemma~\ref{LemmaExpand}}
By changing the indices we get that l.h.s.\ of (\ref{eq2.29}) is, up to a sign, equal to
\begin{equation}\label{eq2.29b}
\det\left[F_{N_i+1-l,k}(x_1^{N_i+1-k}(t_i)-x_1^{l}(t_{i-1}),a,b)\right]_{1\leq k,l\leq N_i}
\end{equation}
Using repeatedly the identity (\ref{eqRecRel1}) we have
\begin{equation}
F_{n,k}(x,a,b)=(\varphi_{N_i+2-k}*\cdots * \varphi_{N_i} * F_{n,1})(x,a,b).
\end{equation}
Therefore,
\begin{equation}
(\ref{eq2.29b})=\det\left[ (\varphi_{N_i+2-k}*\cdots *\varphi_{N_i} * F_{N_i+1-l,1})(x_1^{N_i+1-k}(t_i)-x_1^{l}(t_{i-1}),a,b) \right]_{1\leq i,j \leq N_i}
\end{equation}
We write explicitly the convolution by introducing explicit summation variables as follows
\begin{eqnarray}
& & (\varphi_{N_i+2-k}*\cdots *\varphi_{N_i} * F_{N_i+1-l,1})(x_1^{N_i+1-k}(t_i)-x_1^{l}(t_{i-1}),a,b)\nonumber \\
&=& \sum_{\begin{subarray}{c}x_n^{N_i+1-k+n},\\ 1\leq n \leq k-1\end{subarray}}
 \Big(\prod_{n=1}^{k-1} \varphi_{N_i+1-k+n}(x_n^{N_i-k+n}(t_i),x_{n+1}^{N_i+1-k+n}(t_i))\Big)\nonumber \\
 && \hspace{4em}\times F_{N_i+1-l,1}(x_k^{N_i}(t_i)-x_1^{l}(t_{i-1}),a,b),
\end{eqnarray}
where we used the fact that $\varphi_m(x,y)=\varphi_m(x+c,y+c)$ for any $c\in\Z$. By multi-linearity of the determinant, we can take the sums and the factors $\varphi$'s out of the determinant with the result
\begin{eqnarray}\label{eq4.78}
(\ref{eq2.29b})&=&\sum_{\begin{subarray}{c}x_k^n(t_i),\\ 2\leq k\leq n \leq N_i\end{subarray}}
\Big(\prod_{n=2}^{N_i}\prod_{k=1}^{n-1} \varphi_n(x_k^{n-1}(t_i),x_{k+1}^n(t_i))\Big) \nonumber \\
&\times & \det\left[ F_{N_i+1-l,1}(x_j^{N_i}(t_i)-x_1^{l}(t_{i-1}),a,b) \right]_{1\leq j,l \leq N_i}.
\end{eqnarray}
The product of the $\varphi$'s is non-zero only if $x_k^{n-1}(t_i)\leq x_{k+1}^n(t_i)$. Applying Lemma 3.3 in~\cite{BFPS06} we can further reduce the summation domain to $\widetilde D(t_i)$ without changing the result.

Finally, the product of the determinants of $\varphi$'s in the right-hand side of (\ref{eq2.29}) is either $1$ or $0$ depending on whether the variables interlace (belongs to $D(t_i)$) or not. This implies (\ref{eq2.29}).
\end{proofOF}

\begin{lem}\label{LemmaContract}
We have the identity
\begin{eqnarray}\label{eq2.29c}
& &\sum_{\widehat D(t_i)} \Big(\prod_{n=2}^{N_{i+1}}\det\left[\varphi_n(x_k^{n-1}(t_i),x_l^n(t_i))\right]_{1\leq k,l \leq n}\Big) \nonumber \\
&\times & \det\left[F_{N_{i+1}+1-l,1}(x_k^{N_{i+1}}(t_{i+1})-x_1^{l}(t_i),a,b)\right]_{1\leq k,l\leq N_{i+1}}\nonumber \\
&=& \det\left[F_{1,1}(x_k^{N_{i+1}}(t_{i+1})-x_l^{N_{i+1}}(t_i),a,b)\right]_{1\leq k,l\leq N_{i+1}}.
\end{eqnarray}
\end{lem}
\begin{proofOF}{Lemma~\ref{LemmaContract}}
By an analogue (essentially inverse) procedure as in the proof of Lemma~\ref{LemmaExpand}, we first get
\begin{eqnarray}\label{eq2.29d}
(\ref{eq2.29c})&=&\sum_{\begin{subarray}{c}x_k^n(t_i),\\ 2\leq n \leq N_{i+1}-1,\\ 1\leq k \leq n\end{subarray}}
\Big(\prod_{n=2}^{N_{i+1}}\prod_{k=1}^{n-1} \varphi_n(x_k^{n-1}(t_i),x_{k+1}^n(t_i))\Big) \nonumber \\
&\times &\det\left[F_{N_{i+1}+1-l,1}(x_k^{N_{i+1}}(t_{i+1})-x_1^{l}(t_i),a,b)\right]_{1\leq k,l\leq N_{i+1}}.
\end{eqnarray}
Now we insert by linearity the factor $\prod_{n=l+1}^{N_{i+1}}\varphi_n(x_l^{n-1}(t_i),x_{l+1}^n(t_i))$ to terms $F_{N_{i+1}+1-l,1}(x_k^{N_{i+1}}(t_{i+1})-x_1^l(t_i),a,b)$ as well as the sum over the corresponding variables. The sums are carried out by using (\ref{eqRecRel3}), from which we get the r.h.s.\ of (\ref{eq2.29c}).
\end{proofOF}

\begin{lem}\label{LemPhis}
Let us define
\begin{equation}
\phi_n(x,y) = \varphi_n(x,y),\quad \phi_n(x_n^{n-1},y)=v_n^y.
\end{equation}
Then
\begin{equation}\label{eq4.77}
v_n^{x_1^n}\det\left[\varphi_n(x_k^{n-1},x_l^n)\right]_{1\leq k,l \leq n} =\det\left[\phi_n(x_k^{n-1},x_l^n)\right]_{1\leq k,l \leq n}
\end{equation}
\end{lem}
\begin{proofOF}{Lemma~\ref{LemPhis}}
It is a consequence of the fact that both determinants are zero if the variables $x_i^j$ do not interlace and when they do, the matrices are upper-triangular with diagonal equal to zero and with equal entries in the first $n-1$ rows. The only difference is for the last row, where the matrix in l.h.s.\ of (\ref{eq4.77}) has entries $1$ and r.h.s.\ of (\ref{eq4.77}) has entries $v_n^{x_l^n}$.
\end{proofOF}

\section{Asymptotic analysis}\label{SectAsympt}
\subsection{Flat initial conditions}
To prove Theorem~\ref{ThmFlatCvg} we need the uniform convergence of the kernel in bounded sets as well as bounds uniform in $T$. These results are provided in the following Propositions~\ref{PropBoundedSet},~\ref{PropDiffusion},~\ref{PropModerate}.

Let us define the rescaled and conjugate kernel by
\begin{equation}
K^{\rm resc}_T(u_1,s_1; u_2,s_2) = K((n_1,t_1),x_1; (n_2,t_2),x_2) T^{1/3} \frac{e^{t_2 (2L+R/2)}2^{x_2}}{e^{t_1 (2L+R/2)}2^{x_1}}
\end{equation}
where $n_i=n(u_i)$, $t_i=t(u_i)$, and
\begin{equation}
x_i=[-2 n_i+\textbf{v}\, t_i-s_i T^{1/3}].
\end{equation}

\begin{prop}[Uniform convergence in a bounded set]\label{PropBoundedSet}
Fix $u_1,u_2$, then for any fixed $\ell>0$, the rescaled kernel $K^{\rm resc}_T$ converges uniformly for \mbox{$(s_1,s_2)\in [-\ell,\ell]^2$} as
\begin{equation}
\lim_{T\to\infty}K^{\rm resc}_T(u_1,s_1; u_2,s_2)=S_v^{-1} K_{\Af}(S_h^{-1} u_1,S_v^{-1} s_1;S_h^{-1} u_2,S_v^{-1} s_2),
\end{equation}
with $K_{\Af}$ the kernel of the Airy$_1$ process, see (\ref{eqKernelExpanded}), and $S_v,S_h$ are defined in (\ref{eqS}).
\end{prop}
\begin{proofOF}{Proposition~\ref{PropBoundedSet}}
First we consider the term coming from the second integral in (\ref{eqKernelFlat}), namely
\begin{equation}\label{eq4.27}
\frac{-T^{1/3}}{2\pi\I}\oint_{\Gamma_1}\dx z \frac{e^{R t_1(1-z)+L t_1/(1-z)}}{e^{R t_2 z+L t_2/z}} \frac{z^{n_1+n_2+x_2}}{(1-z)^{n_1+n_2+x_1+1}} \frac{e^{t_2 (2L+R/2)}2^{x_2}}{e^{t_1 (2L+R/2)}2^{x_1}}.
\end{equation}
Define the functions
\begin{eqnarray}\label{eq4.27b}
H(z)&=&Rz+L/z-(R/2-2L)\ln(z),\nonumber\\
g_0(z)&=&(\pi(\theta)+\theta)H(z),\nonumber \\
g_1(z,u)&=&-u (\pi'(\theta)+1)H(z)+u(1-\pi'(\theta))\ln(z(1-z)), \nonumber \\
g_2(z,u,s)&=&u^2 \pi''(\theta) [H(z)+\ln(z(1-z))] +s \ln(z),
\end{eqnarray}
from which we then set
\begin{eqnarray}\label{eq4.27c}
f_0(z)&=&g_0(1-z)-g_0(z),\nonumber \\
f_1(z)&=&g_1(1-z,u_1)-g_1(z,u_2)-g_1(1/2,u_1)+g_1(1/2,u_2),\nonumber \\
f_2(z)&=&g_2(1-z,u_1,s_1)-g_2(z,u_2,s_2)-g_2(1/2,u_1,s_1)+g_2(1/2,u_2,s_2),\nonumber \\
f_3(z)&=&-\ln(1-z).
\end{eqnarray}
With these notations we get
\begin{equation}\label{eq4.28}
(\ref{eq4.27})=\frac{-T^{1/3}}{2\pi\I}\oint_{\Gamma_1}\dx z  e^{T f_0(z)+T^{2/3} f_1(z)+T^{1/3} f_2(z)+f_3(z)}.
\end{equation}
The function $f_0(z)$ has a double critical point at $z=1/2$ and the contribution for large $T$ will be dominated by the one close $z=1/2$. Thus we need to do series expansions around the critical point. Computations leads to
\begin{eqnarray}\label{eq4.31}
f_0(z)&=&\tfrac13 \kappa_0 (z-1/2)^3+\Or((z-1/2)^4), \nonumber \\
f_1(z)&=& -(u_1-u_2)\kappa_1 (z-1/2)^2+\Or((z-1/2)^3),\nonumber \\
f_2(z)&=& -2(s_1+s_2)(z-1/2)+\Or((z-1/2)^2),\nonumber \\
f_3(z)&=&\ln(2)+\Or((z-1/2))
\end{eqnarray}
with
\begin{equation}
\kappa_0=8(8L+R)(\pi(\theta)+\theta),\quad \kappa_1=(R+4L)(\pi'(\theta)+1)+4(1-\pi'(\theta)).
\end{equation}

First we choose $\Gamma_1$ to be a steep descent path\footnote{For an integral $I=\int_\gamma \dx z e^{t f(z)}$, we
say that $\gamma$ is a steep descent path if (1) $\Re(f(z))$ is maximum at some $z_0\in\gamma$: $\Re(f(z))< \Re(f(z_0))$ for $z\in\gamma\setminus\{z_0\}$, and (2) $\Re(f(z))$ is monotone along $\gamma$ except at its maximum point $z_0$ and, if $\gamma$ is closed, at a point $z_1$ where the minimum of $\Re(f)$ is reached.} for $f_0(z)$. Important for the later analysis is that the chosen steep descent path is, close to the critical point, the steepest descent one. We consider $\Gamma_1=\gamma\vee \gamma_c\vee \bar \gamma$, where $\gamma=\{1/2+e^{-\I\pi/3}\xi,0\leq \xi \leq 1/2\}$, $\bar\gamma$ its image with respect to complex conjugation, and $\gamma_c=\{1-1/2 e^{\I\phi},\pi/3\leq \phi\leq 2\pi-\pi/3\}$. We also have $f_0(z)=S_R(z)R(\pi(\theta)+\theta)+S_L(z)L(\pi(\theta)+\theta)$, with
\begin{equation}
S_R(z)=1-2z+\frac12\ln(z/(1-z)),\quad S_L(z)=\frac{1}{1-z}-\frac1z-2\ln(z/(1-z)).
\end{equation}
On $\gamma$, simple computations leads to
\begin{eqnarray}
\Dt{\Re(S_R(z))}{\xi}&=&-\frac{8\xi^2(1+2\xi^2)}{((1+\xi^2)+2\xi^2)((1-\xi)^2+2\xi^2)},\nonumber \\
\Dt{\Re(S_L(z))}{\xi}&=&-\frac{64\xi^2((1+2\xi^2)^2-12\xi^4)}{((1+\xi^2)+2\xi^2)^2((1-\xi)^2+2\xi^2)^2}
\end{eqnarray}
which are both strictly less than $0$ for $\xi\in (0,1/2)$. Now consider the part of $\gamma_c$ with $\phi\in [\pi/3,\pi]$. Then
\begin{eqnarray}\label{eq5.12}
\Dt{\Re(S_R(z))}{\phi}&=&-\frac{4\sin(\phi)(1-\cos(\phi))}{5-4\cos(\phi)},\nonumber \\
\Dt{\Re(S_L(z))}{\phi}&=&-\frac{32\sin(\phi)(1-\cos(\phi))(2-\cos(\phi))}{(5-4\cos(\phi))^2}
\end{eqnarray}
which are both strictly less than $0$ for $\phi \in [\pi/3,\pi)$. For the piece of $\gamma_c$ with $\phi\in(\pi,2\pi-\pi/3]$, (\ref{eq5.12}) is strictly positive, which is right since when $\phi$ increases we go closer to the critical point. Therefore the chosen $\Gamma_1$ is a steep descent path for $f_0(z)$.

Take any $\delta>0$ and set $\Gamma_1^\delta=\{z\in\Gamma_1 | |z-1/2|\leq \delta\}$. Then, if in (\ref{eq4.28}) we integrate only along $\Gamma_1^\delta$ instead of integrating along $\Gamma_1$, the error made is just of order $\Or(e^{-c T})$ for some $c>0$ (more exactly, $c\sim \delta^3$ for $\delta$ small). Thus we now consider the integral on $\Gamma_1^\delta$ only. There, we can use the above series expansions to obtain
\begin{eqnarray}\label{eq4.29}
& &\frac{-2 T^{1/3}}{2\pi\I}\int_{\Gamma_1^\delta}\dx z e^{\frac13 \kappa_0 T (z-1/2)^3+(u_2-u_1)\kappa_1 T^{2/3} (z-1/2)^2-2(s_1+s_2)(z-1/2)}\nonumber \\
& &\times e^{\Or\big(T(z-1/2)^4,T^{2/3}(z-1/2)^3,T^{1/3}(z-1/2)^2,(z-1/2)\big)}.
\end{eqnarray}
The difference between (\ref{eq4.29}) and the same integral without
the error term can be bounded by applying $|e^x-1|\leq |x|e^{|x|}$
to $\Or(\cdots)$. Thus, this error term can be bounded by
\begin{eqnarray}
& &\frac{2 T^{1/3}}{2\pi}\int_{\Gamma_1^\delta}\dx z \Big|e^{\frac13 c_0\kappa_0 T (z-1/2)^3+(u_2-u_1)c_1 \kappa_1 T^{2/3} (z-1/2)^2-2 c_2 (s_1+s_2)(z-1/2)}\nonumber \\
& &\times
\Or\big(T(z-1/2)^4,T^{2/3}(z-1/2)^3,T^{1/3}(z-1/2)^2,(z-1/2)\big)
\Big|
\end{eqnarray}
for some $c_0,c_1,c_2$ which can be taken as close to $1$ as needed by setting $\delta$ small enough. Then, by the change of variable $T^{1/3} (z-1/2)=w$ one gets that this error term is of order $\Or(T^{-1/3})$ (what is needed is just $c_0>0$).

It remains to consider the leading term, namely (\ref{eq4.29}) without the error terms. By extending the integral to infinity by continuing the two small straight segments forming $\Gamma_1^\delta$, the error made is of order $\Or(e^{-c T})$. Thus we obtained that (\ref{eq4.27}) is, up to an error $\Or(e^{-c T},T^{-1/3})$ uniform for \mbox{$s_1,s_2\in [-\ell,\ell]^2$}, equal to
\begin{equation}\label{eq4.30}
\frac{-2 T^{1/3}}{2\pi\I}\int_{\gamma_\infty}\dx z e^{\frac13 \kappa_0 T (z-1/2)^3+(u_2-u_1)\kappa_1 T^{2/3} (z-1/2)^2-2(s_1+s_2)(z-1/2)},
\end{equation}
where $\gamma_\infty$ is a path going from $e^{\I\pi/3}\infty$ to $e^{-\I\pi/3}\infty$.
By the change of variable $w=(\kappa_0 T)^{1/3} (z-1/2)$, we get
\begin{eqnarray}
(\ref{eq4.30})&=&\frac{-1}{2\pi\I}\int_{\gamma_\infty} \dx w \frac{2}{\kappa_0^{1/3}} e^{\frac13 w^3+(u_2-u_1)w^2\kappa_1/\kappa_0^{2/3}-2(s_1+s_2) w/\kappa_0^{1/3}} \\
&=&S_v^{-1} \Ai\big(S_h^{-2} (u_2-u_1)^2+ S_v^{-1} (s_1+s_2)\big)\nonumber \\
&&\times e^{\frac23 S_h^{-3} (u_2-u_1)^3+S_v^{-1} S_h^{-1}(u_2-u_1)(s_1+s_2)}\nonumber
\end{eqnarray}
with $S_v$ and $S_h$ defined in (\ref{eqS}). Here we used the Airy function representation
\begin{equation}
\frac{-1}{2\pi \I}\int_{\gamma_\infty}\dx v
e^{v^3/3+av^2+bv}=\Ai(a^2-b)\exp(2a^3/3-ab).
\end{equation}

To finish the proof, we need to consider the term coming from the first integral in (\ref{eqKernelFlat}), namely
\begin{equation}\label{eq4.40}
-\frac{T^{1/3}}{2\pi\I}\oint_{\Gamma_{0}}\dx w \frac{1}{w^{x_1-x_2+1}} \left(\frac{w}{1-w}\right)^{n_2-n_1} e^{(Rw+L/w)(t_1-t_2)} \frac{e^{t_2 (2L+R/2)}2^{x_2}}{e^{t_1 (2L+R/2)}2^{x_1}}.
\end{equation}
This can be rewritten as
\begin{equation}\label{eq4.41}
(\ref{eq4.40})= \frac{-T^{1/3}}{2\pi\I}\oint_{\Gamma_0}\frac{\dx w}{w} e^{T^{2/3} (p_0(w)-p_0(1/2))+T^{1/3} (p_1(w)-p_1(1/2))}
\end{equation}
with
\begin{eqnarray}
p_0(w)&=&(u_2-u_1)(\pi'(\theta)+1) H(w)-(u_2-u_1)(1-\pi'(\theta))\ln(w(1-w)),\nonumber \\
p_1(w)&=&-(u_2^2-u_1^2)\frac{\pi''(\theta)}{2} [H(w)+\ln(w(1-w))]-(s_2-s_1)\ln(w),
\end{eqnarray}
where $H(w)$ is the function defined in (\ref{eq4.27b}). Remark that we need to do the analysis only for $u_2>u_1$. The function $p_0$ has critical point at $w=1/2$. The series expansions of $p_0$ and $p_1$ around $w=1/2$ are
\begin{eqnarray}
p_0(w)&=&p_0(1/2)+\kappa_1 (u_2-u_1)(w-1/2)^2+\Or((w-1/2)^3),\nonumber \\
p_1(w)&=&p_1(1/2)+2(s_1-s_2)(w-1/2)+\Or((w-1/2)^2).
\end{eqnarray}
We choose as path $\Gamma_0=\{\tfrac12 e^{\I \phi},\phi\in (-\pi,\pi]\}$. This is a steep descent path for $p_0$. In fact, for $w\in\Gamma_0$,
\begin{eqnarray}
\Re(H(w))&=&(R/2+2L)\cos(\phi)+(R/2-2L)\ln(2), \\
\Re(-\ln(w(1-w)))&=&\ln(2)-\ln|1-w|=2\ln(2)-\tfrac12 \ln(5-4\cos(\phi)),\nonumber
\end{eqnarray}
which are decreasing when $\cos(\phi)$ decreases. Thus, we can integrate only on $\Gamma_0^\delta=\{w\in\Gamma_0| |w-1/2|\leq\delta\}$ and, for a small $\delta$, the error term is just of order $\Or(e^{-c T^{2/3}})$ with $c>0$ ($c\sim \delta^2$ as $\delta\ll 1$). The integral over $\Gamma_0^\delta$ is then given by
\begin{eqnarray}\label{eq4.45}
& &\frac{-2T^{1/3}}{2\pi\I}\int_{\Gamma_0^\delta}\dx w
e^{\kappa_1 (u_2-u_1)(w-1/2)^2T^{2/3}+2(s_1-s_2)(w-1/2)T^{1/3}} \nonumber \\
& &\times e^{\Or((w-1/2)^3T^{2/3},(w-1/2)^2T^{1/3},(w-1/2))}.
\end{eqnarray}
As above, we use $|e^{x}-1|\leq |x|e^{|x|}$, to control the difference between (\ref{eq4.45}) and the same expression without the error terms. By taking $\delta\ll 1$ and the change of variable $\tilde w=(w-1/2)T^{1/3}$, we get that this difference is of order $\Or(T^{-1/3})$ uniformly for $s_1,s_2$ in a bounded set. Once we have taken away the error terms in (\ref{eq4.45}), we extend the integral to $1/2\pm \I\infty$. By this we make only an error of order $\Or(e^{-c T^{2/3}})$. The integration path can be deformed to $1/2+\I\R$ without passing through any poles, therefore by setting $w=1/2+\I y T^{-1/3}$ we get
\begin{eqnarray}
& &-\frac{1}{\pi}\int_{\R}\dx y e^{-\kappa_1(u_2-u_1) y^2+2(s_1-s_2)} =-\frac{1}{\sqrt{\pi\kappa_1(u_2-u_1)}}\exp\left(-\frac{(s_2-s_1)^2}{\kappa_1(u_2-u_1)}\right)\nonumber \\
&=&-\frac{S_v^{-1}}{\sqrt{4\pi(u_2-u_1)S_h^{-1}}}\exp\left(-\frac{(s_2-s_1)^2 S_v^{-2}}{4 (u_2-u_1)S_h^{-1}}\right).
\end{eqnarray}
Since all the error terms in the series expansions are uniform for \mbox{$(s_1,s_2)\in[-\ell,\ell]^2$}, the result of the Proposition is proven.
\end{proofOF}

\begin{prop}[Bound for the diffusion term of the kernel]\label{PropDiffusion}
$ $ \\ For any $s_1,s_2\in\R$ and $u_2-u_1>0$ fixed, the bound
\begin{eqnarray}\label{eq4.47}
& &\bigg| \frac{e^{t_2 (2L+R/2)}2^{x_2}}{e^{t_1 (2L+R/2)}2^{x_1}}
\frac{T^{1/3}}{2\pi\I} \oint_{\Gamma_{0}}\dx w \frac{1}{w^{x_1-x_2+1}} \left(\frac{w}{1-w}\right)^{n_2-n_1} e^{(Rw+L/w)(t_1-t_2)}\bigg| \nonumber \\
& &\leq \cte \, e^{-|s_1-s_2|}
\end{eqnarray}
holds for $T$ large enough and $\cte$ independent of $T$. The $\cte$ is uniform in $s_1,s_2$ but not in $u_2-u_1$.
\end{prop}
\begin{proofOF}{Proposition~\ref{PropDiffusion}}
From the analysis in Proposition~\ref{PropBoundedSet}, we just need a bound for $|s_2-s_1|\geq \ell$, $\ell>0$ fixed. We start with (\ref{eq4.41}) but to obtain a decaying bound for large $|s_2-s_1|$ we consider another path $\Gamma_0$.

Consider an $\e$ with $0<\e \ll 1$ and set $\Gamma_0=\{w=\rho e^{\I\phi},\phi\in[-\pi,\pi)\}$, with
\begin{equation}\label{eq4.48}
\rho=\left\{\begin{array}{ll}
\frac12 + \frac{(s_2-s_1)T^{-1/3}}{(u_2-u_1)\kappa_1},&\textrm{if }|s_2-s_1|\leq \e T^{1/3},\\
\frac12 + \frac{\e}{(u_2-u_1)\kappa_1},&\textrm{if }\,\,s_2-s_1\,\, \geq \e T^{1/3},\\
\frac12 - \frac{\e}{(u_2-u_1)\kappa_1},&\textrm{if }\,\,s_2-s_1\,\, \leq - \e T^{1/3}.
\end{array}\right.
\end{equation}
We have $\Dt{ }{\phi} \Re(w-\tfrac12\ln(w))=-\rho \sin(\phi)$,
$\Dt{ }{\phi}\Re(1/w+2\ln(w))=-\frac{4}{\rho}\sin(\phi)$, and $\Dt{ }{\phi}\Re(-\ln(w(1-w)))=-\frac{\rho\sin(\phi)}{1-2\rho\cos(\phi)+\rho^2}$. Thus $\Gamma_0$ is a steep descent path for $p_0(w)$. Moreover, since on $\Gamma_0$ we have $\Re(\ln(w))=\ln(\rho)$ is a constant, $\Gamma_0$ is also a steep descent path for $p_0(w)$ plus the term of $p_1(w)$ proportional to $s_2-s_1$. Let, for a small $\delta>0$ fixed, $\Gamma_0^\delta=\{w=\rho e^{\I\phi},\phi\in(-\delta,\delta)\}$. Then
\begin{eqnarray}\label{eq4.49}
(\ref{eq4.41}) &=& e^{T^{2/3} (p_0(\rho)-p_0(1/2))+T^{1/3} (p_1(\rho)-p_1(1/2))} \\
& \times & \bigg(\Or(e^{-c T^{2/3}})+\frac{-T^{1/3}}{2\pi\I}\int_{\Gamma_0^\delta}\frac{\dx w}{w} e^{T^{2/3} (p_0(w)-p_0(\rho))+T^{1/3} (p_1(w)-p_1(\rho))}\bigg)\nonumber
\end{eqnarray}
for some $c>0$ (for small $\delta$, $c\sim \delta^2$). On $\Gamma_0^\delta$ the $s_i$-dependent term in $\Re(p_1(w)-p_1(\rho))$ is equal to zero and the rest is of order $\Or(\phi^2)$. Therefore the last integral can be bounded by
\begin{equation}\label{eq4.50}
\frac{T^{1/3}}{2\pi}\int_{-\delta}^\delta \frac{\dx \phi}{\rho} e^{-\tfrac12 T^{2/3} (u_2-u_1) \big[ (\pi'(\theta)+1)(R\rho+L/\rho)+(1-\pi'(\theta))\rho/(1-\rho)^2\big]\phi^2+\Or(T^{2/3}\phi^4,T^{1/3}\phi^2)}.
\end{equation}
For $\delta$ small enough, and $T$ large enough, the terms $\Or(T^{2/3}\phi^4)$ and $\Or(T^{1/3}\phi^2)$ are both controlled by the first term in the exponential. Then, by the change of variable $T^{1/3}\phi=\psi$ one sees that r.h.s.\ of (\ref{eq4.50}) is bounded by a constant, uniformly in $T$.

What remains is therefore to bound the first term in the r.h.s.\ of (\ref{eq4.49}). By the choice in (\ref{eq4.48}) of $\rho$, $|\rho-1/2|\leq \e/((u_2-u_1)\kappa_1) \ll 1$ for $\e$ small enough which can be still chosen. Series expansion for $\rho$ close to $1/2$ leads to
\begin{eqnarray}
p_0(\rho)-p_0(1/2)&=&-2 (s_2-s_1) (\rho-1/2) T^{1/3}(1+\Or(\rho-1/2))\nonumber \\
&+& \kappa_1 (u_2-u_1) (\rho-1/2)^2T^{2/3}(1+\Or(\rho-1/2)).
\end{eqnarray}
By (\ref{eq4.48}) we obtain the bounds
\begin{eqnarray}
p_0(\rho)-p_0(1/2)&=&-\frac{(s_2-s_1)^2}{(u_2-u_1)\kappa_1}(1+\Or(\e)),\textrm{ if }|s_2-s_1|\leq \e T^{1/3}, \\
p_0(\rho)-p_0(1/2)&=&-\frac{|s_2-s_1|\e T^{1/3}}{(u_2-u_1)\kappa_1}(1+\Or(\e)),\textrm{ if }|s_2-s_1|\geq \e T^{1/3}.\nonumber
\end{eqnarray}
Combining the above result we have
\begin{equation}\label{eq4.53}
|(\ref{eq4.47})|\leq \Big[\Or(e^{-c T^{2/3}})+ \Or(1)\Big] \Big[e^{-\frac{(s_2-s_1)^2}{(u_2-u_1)\kappa_1}(1+\Or(\e))} + e^{-\frac{|s_2-s_1|\e T^{1/3}}{(u_2-u_1)\kappa_1}(1+\Or(\e))}\Big].
\end{equation}
Thus by taking an $\e$ small enough and then $T$ large enough the bound (\ref{eq4.53}) implies the statement to be proven, since for any $\alpha>0$, there exists a $C_\alpha<\infty$ such that $e^{-\alpha (s_2-s_1)^2}\leq C_\alpha e^{-|s_2-s_1|}$.
\end{proofOF}

\begin{prop}[Bound on the main term of the kernel]\label{PropModerate}
$ $ \\Let $u_1,u_2$ be fixed. Then, for any $(s_1,s_2)\in [-\ell,\infty)^2$, the bound
\begin{eqnarray}
& &\left|\frac{-T^{1/3}}{2\pi\I}\oint_{\Gamma_1}\dx z \frac{e^{R t_1(1-z)+L t_1/(1-z)}}{e^{R t_2 z+L t_2/z}} \frac{z^{n_1+n_2+x_2}}{(1-z)^{n_1+n_2+x_1+1}} \frac{e^{t_2 (2L+R/2)}2^{x_2}}{e^{t_1 (2L+R/2)}2^{x_1}}\right| \nonumber \\
& &\leq \cte \, e^{-(s_1+s_2)}
\end{eqnarray}
holds for $T$ large enough, where $\cte$ is a constant independent of $T$.
\end{prop}
\begin{proofOF}{Proposition~\ref{PropModerate}}
Let $\tilde\ell$ be a constant independent from $T$, which can still be chosen large if needed. For $(s_1,s_2)\in [-\ell,\tilde\ell]^2$, the result is a consequence of the estimates in the proof of Proposition~\ref{PropBoundedSet}. Therefore we can consider just $(s_1,s_2)\in [-\ell,\infty)^2\setminus [-\ell,\tilde\ell]^2$. Introduce the notation \mbox{$\tilde s_i=(s_i+\ell+\tilde\ell) T^{-2/3}$}, which then belongs to $[\tilde\ell T^{-2/3},\infty)$.

The integral to be bounded is
\begin{equation}\label{eq4.54}
\frac{-T^{1/3}}{2\pi\I}\oint_{\Gamma_1}\dx z  e^{T f_0(z)+T^{2/3} f_1(z)+T^{1/3} f_2(z)+f_3(z)}
\end{equation}
where $f_1(z)$ and $f_3(z)$ are given in (\ref{eq4.27c}), and $f_0(z)$ and $f_2(z)$ are just slight modifications of the functions in (\ref{eq4.27c}), namely
\begin{eqnarray}
f_0(z)&=& (\pi(\theta)+\theta) (H(1-z)-H(z))+\tilde s_1 \ln(2(1-z))-\tilde s_2\ln(2z),\nonumber \\
f_2(z)&=&g_2(1-z,u_1,-\ell-\tilde\ell)-g_2(z,u_2,-\ell-\tilde\ell)\\
& &-g_2(1/2,u_1,-\ell-\tilde\ell)+g_2(1/2,u_2,-\ell-\tilde\ell).\nonumber
\end{eqnarray}
We put $\tilde s_1$ and $\tilde s_2$ in $f_0(z)$, because they are not restricted to be of order $T^{-2/3}$ (as it was the case in Proposition~\ref{PropBoundedSet}).

First we need to find a steep descent path for $f_0(z)$. We choose it as $\Gamma_1=\{1-\rho e^{\I\phi},\phi\in[-\pi,\pi)\}$ with $0<\rho\leq 1/2$, chosen as follows,
\begin{eqnarray}\label{eq4.55}
\rho=\left\{\begin{array}{ll}
\frac12-((\tilde s_1+\tilde s_2)/\kappa_0)^{1/2},& \tilde s_1+\tilde s_2\leq \e,\\
\frac12-(\e/\kappa_0)^{1/2},& \tilde s_1+\tilde s_2\geq \e,
\end{array}\right.
\end{eqnarray}
for some small $\e>0$ to be fixed later. Recall that the $\tilde s_i>0$.

To see that $\Gamma_1$ is a steep descent path, we consider $f_0(z)$ term by term. First consider $\phi\in [0,\pi]$, the case $\phi\in[-\pi,0]$ is obtained by symmetry. The term proportional to $R(\pi(\theta)+\theta)$ satisfies
\begin{equation}
\Dt{ }{\phi}\Re(1-2z+\tfrac12\ln(z/(1-z))) = -\frac{\rho(3-8\rho\cos(\phi)+4\rho^2)\sin(\phi)}{1-2\rho\cos(\phi)+\rho^2}\leq 0
\end{equation}
for all $0<\rho\leq 1/2$, with equality only at $\phi=0,\pi$. The term proportional to $L(\pi(\theta)+\theta)$ satisfies
\begin{equation}
\Dt{ }{\phi}\Re(1/(1-z)-1/z-2\ln(z/(1-z))) = -\frac{((1-2\rho\cos(\phi)+2\rho^2)^2-\rho^2)\sin(\phi)}{(1-2\rho\cos(\phi)+\rho^2)^2\rho}\leq 0
\end{equation}
for all $0<\rho\leq 1/2$, with equality only at $\phi=0,\pi$. Finally, $\Re(\ln(1-z))$ is constant on $\Gamma_1$ and $-\Re(\ln(2z))=-\ln(2|z|)$ is strictly decreasing while moving on $\Gamma_1$ with $|\phi|$ increasing.

For a small $\delta>0$, $\Gamma_1^{\delta}=\{1-\rho e^{\I\phi},\phi\in (-\delta,\delta)\}$. We also define
\begin{equation}
Q(\rho)=\exp\left(\Re\big(T f_0(1-\rho)+T^{2/3}f_1(1-\rho)+T^{1/3}f_2(1-\rho)\big)\right).
\end{equation}
Since $\Gamma_1$ is a steep descent path of $f_0(z)$, the integral over $\Gamma_1\setminus\Gamma_1^\delta$ is bounded by
\begin{equation}\label{eq4.59}
Q(\rho) \Or(e^{-c T})
\end{equation}
for some $c>0$ independent of $T$. The contribution of the integral over $\Gamma_1^\delta$ is bounded by
\begin{equation}\label{eq4.60}
Q(\rho) \bigg|\frac{-T^{1/3}}{2\pi\I}\int_{\Gamma_1^\delta} \dx z e^{T(f_0(z)-f_0(1-\rho))+T^{2/3}(f_1(z)-f_1(1-\rho))+T^{1/3}(f_2(z)-f_2(1-\rho))+f_3(z)}\bigg|
\end{equation}
The series expansion around $\phi=0$ is
\begin{equation}
\Re(f_0(1-\rho e^{\I\phi})-f_0(1-\rho))=-\gamma_1 \phi^2(1+\Or(\phi))
\end{equation}
with
\begin{equation}
\gamma_1=\frac{\tilde s_2\rho}{2(1-\rho)^2}+\frac{(\pi(\theta)+\theta)(1-2\rho)}{(1-\rho)^2} \left(\frac{R\rho(3-2\rho)}{4}+\frac{L(1-\rho+2\rho^2)}{3\rho(1-\rho)}\right),
\end{equation}
and
\begin{equation}
\Re(f_1(1-\rho e^{\I\phi})-f_1(1-\rho))=\gamma_2 \phi^2 (1+\Or(\phi)),
\end{equation}
with
\begin{equation}
\gamma_2= (u_2-u_1)\kappa_1 + \Or(\rho-1/2).
\end{equation}
Finally, $\Re(f_2(1-\rho e^{\I\phi})-f_2(1-\rho))=\Or(\phi^2)$.
Thus, by the change of variable $z=1-\rho e^{\I\phi}$, the above estimates, and by setting $\gamma=\gamma_1+\gamma_2 T^{-1/3}$, we get
\begin{equation}
(\ref{eq4.60}) = Q(\rho) \frac{T^{1/3}\rho}{2\pi(1-\rho)}\int_{-\delta}^\delta \dx \phi e^{-\gamma \phi^2 T(1+\Or(\phi))(1+\Or(T^{-1/3}))}.
\end{equation}
By choosing $\delta$ small enough (independent of $T$) and then $T$
large enough, the factors with the error terms can be replaced by
$1/2$, thus
\begin{equation}
(\ref{eq4.60}) \leq Q(\rho) \frac{T^{1/3}\rho}{2\pi(1-\rho)}\int_{-\delta}^\delta \dx \phi e^{-\gamma \phi^2 T/2} \leq Q(\rho)\frac{1}{\sqrt{2\pi\gamma T^{1/3}}}.
\end{equation}
Remark that, the worse case is when $\gamma$ becomes small, and this
happens when $\rho\to 1/2$, i.e., it is the case of small values of
$\tilde s_1+\tilde s_2$. But even in this case, $\gamma_1
T^{1/3}\sim (s_1+s_2+2\ell+2\tilde\ell)^{1/2}\geq
(2\tilde\ell)^{1/2}$. Since $\gamma_2$ is of order one, $\gamma T^{1/3}=\gamma_1 T^{1/3}+\gamma_2>0$ for $\tilde\ell$ large enough. So, for by setting $\tilde\ell$ large enough, $(\ref{eq4.60})\leq \cte Q(\rho)$. This
estimate, combined with (\ref{eq4.59}), implies that the Proposition
will be proven by showing that $Q(\rho)\leq \cte e^{-(s_1+s_2)}$.
Since $1-\rho$ is close to $1/2$, we can apply the series expansion
of $f_i$ around $z=1/2$. The expansion of $f_1$ is in
(\ref{eq4.31}), while the one of $f_2$ is the same as in
(\ref{eq4.31}) with $s_1+s_2=-2\ell-2\tilde\ell$. Finally,
\begin{equation}
f_0(z)=\tfrac13\kappa_0 (z-1/2)^3 (1+\Or(z-1/2)^2)-(\tilde s_1+\tilde s_2)(z-1/2)(1+\Or(z-1/2)).
\end{equation}
First consider $\tilde s_1+\tilde s_2\leq \e$. Then, with $\rho$ chosen as in (\ref{eq4.55}), we get
\begin{eqnarray}
Q(\rho)&=&e^{-\tfrac23 T (\tilde s_1+\tilde s_2)^{3/2}\kappa_0^{-1/2}T(1+\Or(\sqrt{\e}))}
e^{(u_2-u_1)\kappa_1(\tilde s_1+\tilde s_2) T^{2/3}\kappa_0^{-1}(1+\Or(\sqrt{\e}))}\nonumber \\
& &\times e^{-2(\ell+\tilde\ell)(\tilde s_1+\tilde s_2) \kappa_0^{-1/2} T^{1/3} (1+\Or(\sqrt{\e}))}\nonumber \\
&=& e^{-\tfrac23 (s_1+s_2+2\ell+2\tilde\ell)^{3/2}\kappa_0^{-1/2}(1+\Or(\sqrt{\e}))}
e^{(u_2-u_1)\kappa_1(s_1+s_2+2\ell+2\tilde\ell)\kappa_0^{-1}(1+\Or(\sqrt{\e}))}\nonumber \\
& & \times e^{-2(\ell+\tilde\ell)(s_1+s_2+2\ell+2\tilde\ell) \kappa_0^{-1/2} T^{-1/3} (1+\Or(\sqrt{\e}))}.
\end{eqnarray}
Recall that $s_1+s_2+2\ell+2\tilde\ell\geq 2\tilde\ell \gg 1$ for $\tilde\ell\gg 1$. Therefore by choosing $\tilde\ell$ large enough (depending only on the coefficients $\kappa_0,\kappa_1,u_1,u_2$ which are however fixed), all the terms are controlled by the first one, i.e.,
\begin{equation}
Q(\rho)\leq e^{-\tfrac13 (s_1+s_2+2\ell+2\tilde\ell)^{3/2}\kappa_0^{-1/2}}\leq e^{-\tfrac13 (s_1+s_2)^{3/2}\kappa_0^{-1/2}}.
\end{equation}
Since this decays more rapidly that $\exp(-(s_1+s_2))$, the Proposition holds for $\tilde s_1+\tilde s_2\leq \e$.

The last case is $\tilde s_1+\tilde s_2\geq \e$. In this case, with $\rho$ chosen as in (\ref{eq4.55}), we obtain
\begin{eqnarray}
Q(\rho)&=&e^{T\kappa_0^{-1/2}(1+\Or(\sqrt{\e}))\sqrt{\e}(\e/3-(\tilde s_1+\tilde s_2))}
e^{(u_2-u_1)\kappa_1 \kappa_0^{-1} \e T^{2/3}(1+\Or(\sqrt{\e}))}\nonumber \\
& &\times e^{-4(\ell+\tilde\ell) \kappa_0^{-1/2} \e T^{1/3}(1+\Or(\sqrt{\e}))}.
\end{eqnarray}
But now, $\e/3-(\tilde s_1+\tilde s_2)\leq -\tfrac23 (\tilde s_1+\tilde s_2)$, thus the first term in the exponential is, up to a positive constant,$-\sqrt{\e}T^{1/3}(s_1+s_2+2\ell+2\tilde\ell)$, which dominates the second term $\sim \e T^{2/3}\leq s_1+s_2+2\ell+2\tilde\ell$, and it also dominates the third term. Therefore, for any choice of $\e$ and $\tilde\ell$ made before, we can take $T$ large enough such that
\begin{equation}
Q(\rho)\leq e^{-\frac13 \sqrt{\e} T^{1/3}(s_1+s_2)},
\end{equation}
which ends the proof of the Proposition.
\end{proofOF}

\begin{proofOF}{Theorem~\ref{ThmFlatCvg}}
The proof of Theorem~\ref{ThmFlatCvg} is the complete analogue of Theorem 2.5 in~\cite{BFP06}. The results in Propositions 5.1,5.3,5.4, and 5.5 in~\cite{BFP06} are replaced by the ones in Proposition~\ref{PropBoundedSet},~\ref{PropDiffusion},~\ref{PropModerate}. The strategy is to write the Fredholm series of the expression for finite $T$ and, by using the bounds in Propositions~\ref{PropDiffusion} and~\ref{PropModerate}, see that it is bounded by a $T$-independent and integrable function. Once this is proven, one can exchange the sums/integrals and the $T\to\infty$ limit by the theorem of dominated convergence. For details, see Theorem 2.5 in~\cite{BFP06}.
\end{proofOF}

\subsection{Sketch of the result (\ref{eq4.11})}\label{SectStepIC}
With the rescaling (\ref{eqScaling1}) and (\ref{eqScaling2}), the rescaled kernel writes
\begin{equation}
K^{\rm resc}(u_1,s_1; u_2,s_2) = K((n_1,t_1),x_1; (n_2,t_2),x_2) T^{1/3}.
\end{equation}
The main part of the kernel (the second term in (\ref{eqKernelStep})) writes
\begin{equation}\label{eq4.10}
\frac{T^{1/3}}{(2\pi\I)^2}\oint_{\Gamma_0}\dx w \oint_{\Gamma_1}\dx z \frac{e^{T f_0(w)+T^{2/3} f_1(w;u_1)+T^{1/3} f_2(w;u_1,s_1)}}{e^{T f_0(z)+T^{2/3} f_1(z;u_2)+T^{1/3} f_2(z;u_2,s_2)}}\frac{1}{w(w-z)}
\end{equation}
with
\begin{eqnarray}
& &f_0(w)=(\pi(\theta)+\theta)\left(R w + \tfrac{L}{w}\right)+(\pi(\theta)-\theta)\ln\left(\tfrac{1-w}{w}\right)-\sigma_0 \ln(w),\nonumber \\
& &f_1(w;u_i)=-\left[(\pi'(\theta)+1)\left(R w + \tfrac{L}{w}\right)+(\pi'(\theta)-1)\ln\left(\tfrac{1-w}{w}\right)-\sigma_1\ln(w)\right] u_i,\nonumber\\
& &f_2(w;u_i,s_i)=\left[\tfrac12 \pi''(\theta)\left(R w + \tfrac{L}{w}+\ln\left(\tfrac{1-w}{w}\right)\right)-\sigma_2\right]u_i^2+s_i \ln(w).
\end{eqnarray}

The parameter $\mu$ is actually the position of the double critical point of $f_0(w)$. Series expansions gives
\begin{eqnarray}
f_0(w)&=& f_0(\mu)-\frac{\kappa_0}{3}(w-\mu)^3+\Or((w-\mu)^4),\nonumber \\
f_1(w;u_1)&=&f_1(\mu;u_1)-u_1 \kappa_1 (w-\mu)^2 + \Or((w-\mu)^3),\\
f_2(w;u_1,s_1)&=&f_2(\mu;u_1,s_1)-\left(\frac{\kappa_1^2 u_1^2}{\kappa_0}-\frac{s_1}{\mu}\right)(w-\mu)+\Or((w-\mu)^2). \nonumber
\end{eqnarray}
The terms $f_1(\mu;u_i)$ and $f_2(\mu;u_i,s_i)$ cancel out by an appropriate conjugation of the kernel (\ref{eq4.10}). We denote by $\simeq$ an equality up to conjugation. Thus, asymptotically, (\ref{eq4.10}) goes to
\begin{equation}\label{eq4.14}
\frac{T^{1/3}}{\mu (2\pi\I)^2}\oint_{\Gamma_0}\dx w \oint_{\Gamma_1}\frac{\dx z}{w-z} \frac{e^{-\kappa_0(w-\mu)^3 T/3-u_1 \kappa_1 (w-\mu)^2 T^{2/3}+ T^{1/3}(w-\mu) (s_1/\mu-\kappa_1^2 u_1^2/\kappa_0)}}
{e^{-\kappa_0(z-\mu)^3 T/3-u_2 \kappa_1 (z-\mu)^2 T^{2/3}+T^{1/3}(z-\mu) (s_2/\mu-\kappa_1^2 u_2^2/\kappa_0)}}
\end{equation}
With the change of variable $(w-\mu) (\kappa_0 T)^{1/3}=W$, $(z-\mu) (\kappa_0 T)^{1/3}=Z$, we then obtain
\begin{equation}\label{eq4.16}
(\ref{eq4.14}) = \frac{\kappa_0^{-1/3}}{\mu(2\pi\I)^2}\int \dx W \int \dx Z \frac{1}{W-Z}
\frac{e^{\frac13 Z^3+u_2 Z^2 \kappa_1/\kappa_0^{2/3}-Z(s_2/\mu-\kappa_1^2 u_2^2/\kappa_0)/\kappa_0^{1/3}}}
{e^{\frac13 W^3+u_1 W^2 \kappa_1/\kappa_0^{2/3}-W(s_1/\mu-\kappa_1^2 u_1^2/\kappa_0)/\kappa_0^{1/3}}}.
\end{equation}
Let us denote by $\tilde S_v=\mu\kappa_0^{1/3}$ and $\tilde S_h=\kappa_1^{-1} \kappa_0^{2/3}$ the vertical and horizontal scaling. Then
\begin{equation}
(\ref{eq4.16})=\tilde S_v^{-1} K_{\Ac}(\tilde S_h^{-1} u_1,\tilde S_v^{-1} s_1;\tilde S_h^{-1} u_2,\tilde S_v^{-1} s_2)
\end{equation}
where $K_{\Ac}$ is the extended Airy kernel associated to the Airy$_2$ process. An asymptotic analysis of large deviations similar to Propositions~\ref{PropDiffusion} and~\ref{PropModerate} above would then lead to the result of (\ref{eq4.11}).

\end{document}